\documentclass[preprint2,flushrt]{aastex}

\usepackage{psfig}
\usepackage{epsfig}

\def\mbar{\ifmmode\overline{m}\else$\overline{m}$\fi}
\def\Mbar{\ifmmode\overline{M}\else$\overline{M}$\fi}
\def\mibar{\ifmmode\overline{m}_I\else$\overline{m}_I$\fi}
\def\MIbar{\ifmmode\overline{M}_I\else$\overline{M}_I$\fi}
\def\Nbar{\ifmmode\overline{N}\else$\overline{N}$\fi}

\def\ho{\ifmmode H_0\else$H_0$\fi}

\def\dmod{\ifmmode(m{-}M)_0\else$(m{-}M)_0$\fi}
\def\mM{\ifmmode(m{-}M)_0\else$(m{-}M)_0$\fi}
\def\vi{\ifmmode(V{-}I)\else$(V{-}I)$\fi}
\def\viz{\ifmmode(V{-}I)_0\else$(V{-}I)_0$\fi}
\def\EBV{\ifmmode E_{B-V}\else$E_{B-V}$\fi}

\begin{document}

\title{23 High Redshift Supernovae from the IfA Deep
Survey\altaffilmark{1,2}: Doubling the SN Sample at $z>0.7$}

\author{Brian J. Barris\altaffilmark{3}, John
L. Tonry\altaffilmark{3}, St\'{e}phane Blondin\altaffilmark{4}, Peter
Challis\altaffilmark{5}, Ryan Chornock\altaffilmark{6}, Alejandro
Clocchiatti\altaffilmark{7}, Alexei V. Filippenko\altaffilmark{6},
Peter Garnavich\altaffilmark{8}, Stephen T. Holland\altaffilmark{9},
Saurabh Jha\altaffilmark{6}, Robert P. Kirshner\altaffilmark{5}, Kevin
Krisciunas\altaffilmark{10,11}, Bruno Leibundgut\altaffilmark{4},
Weidong Li\altaffilmark{6}, Thomas Matheson\altaffilmark{5}, Gajus
Miknaitis\altaffilmark{12}, Adam G. Riess\altaffilmark{13}, Brian
P. Schmidt\altaffilmark{14}, R. Chris Smith\altaffilmark{10,11},
Jesper Sollerman\altaffilmark{15}, Jason Spyromilio\altaffilmark{4},
Christopher W. Stubbs\altaffilmark{12}, Nicholas
B. Suntzeff\altaffilmark{10,11}, Herv\'{e} Aussel\altaffilmark{3},
K. C. Chambers\altaffilmark{3}, M. S. Connelley\altaffilmark{3},
D. Donovan\altaffilmark{3}, J. Patrick Henry\altaffilmark{3}, Nick
Kaiser\altaffilmark{3}, Michael C. Liu\altaffilmark{3,16}, Eduardo
L. Mart\'\i n\altaffilmark{3}, and Richard
J. Wainscoat\altaffilmark{3} }

\altaffiltext{1}{CFHT: Based in part on observations obtained at the
Canada-France-Hawaii Telescope (CFHT) which is operated by the 
National Research Council of Canada, the Institut National des 
Science de l'Univers of the Centre National de la Recherche 
Scientifique of France, and the University of Hawaii.  
CTIO: Based in part on observations taken at the Cerro-Tololo
Inter-American Observatory.
Keck: Some of the data presented herein were obtained at the W.M. Keck
Observatory, which is operated as a scientific partnership among the
California Institute of Technology, the University of California, and
the National Aeronautics and Space Administration.  The Observatory
was made possible by the generous financial support of the W.M. Keck
Foundation.  
Magellan: Based in part on observations from the 6.5 m Baade telescope
operated by the Observatories of the Carnegie Institution of
Washington for the Magellan Consortium, a collaboration between the
Carnegie Observatories, the University of Arizona, Harvard University,
the University of Michigan, and the Massachusetts Institute of
Technology.
UH: Based in part on observations with the University of
Hawaii 2.2-m telescope at Mauna Kea Observatory, Institute for
Astronomy, University of Hawaii.
VLT: Based in part on observations obtained at the European Southern
Observatory, Paranal, Chile, under programs ESO 68.A-0427.
}
\altaffiltext{2}{Based in part on observations with the NASA/ESA
$Hubble\ Space\ Telescope$, obtained at the Space Telescope Science
Institute, which is operated by the Association of Universities for
Research in Astronomy (AURA), Inc., under NASA contract NAS 5-26555.
This research is primarily associated with proposal GO-09118.}

\altaffiltext{3}{Institute for Astronomy (IfA), University of Hawaii,
2680 Woodlawn Drive, Honolulu, HI 96822; barris@ifa.hawaii.edu,
jt@ifa.hawaii.edu, aussel@ifa.hawaii.edu, chambers@ifa.hawaii.edu,
msc@galileo.ifa.hawaii.edu, donovan@ifa.hawaii.edu,
henry@ifa.hawaii.edu, kaiser@ifa.hawaii.edu, liu@ifa.hawaii.edu,
ege@ifa.hawaii.edu, rjw@ifa.hawaii.edu}

\altaffiltext{4}{European Southern Observatory,
Karl-Schwarzschild-Strasse 2, Garching, D-85748,
Germany; sblondin@eso.org, bleibund@eso.org, jspyromi@eso.org}

\altaffiltext{5}{Harvard-Smithsonian Center for Astrophysics, 60
Garden Street, Cambridge, MA 02138; pchallis@cfa.harvard.edu,
kirshner@cfa.harvard.edu, tmatheson@cfa.harvard.edu}

\altaffiltext{6}{University of California, Department of
Astronomy, 601 Campbell Hall, Berkeley, CA
94720-3411; rchornock@astro.berkeley.edu, alex@astro.berkeley.edu,
sjha@astro.berkeley.edu, weidong@astro.berkeley.edu}

\altaffiltext{7}{Pontificia Universidad Cat\'{o}lica de Chile,
Departmento de Astronom\'\i a y Astrof\'\i sica, Casilla 306, Santiago
22, Chile; aclocchi@astro.puc.cl.}

\altaffiltext{8}{University of Notre Dame, Department of Physics, 225
Nieuwland Science Hall, Notre Dame, IN 46556-5670; pgarnavi@miranda.phys.nd.edu}

\altaffiltext{9}{Code 662.20, Goddard Space Flight Centre, Greenbelt,
MD 20771-0003; sholland@milkyway.gsfc.nasa.gov}

\altaffiltext{10}{Cerro Tololo Inter-American Observatory, Casilla
603, La Serena, Chile; kkrisciunas@noao.edu, csmith@ctio.noao.edu,
nsuntzeff@noao.edu}

\altaffiltext{11}{Las Campanas Observatory, Casilla 601, La Serena, Chile}

\altaffiltext{12}{University of Washington, Department of Astronomy,
Box 351580, Seattle, WA 98195-1580; gm@u.washington.edu,
stubbs@astro.washington.edu}

\altaffiltext{13}{Space Telescope Science Institute, 3700 San Martin
Drive, Baltimore, MD 21218; ariess@stsci.edu}

\altaffiltext{14}{The Research School of Astronomy and Astrophysics,
The Australian National University, Mount Stromlo and Siding Spring
Observatories, via Cotter Rd, Weston Creek PO 2611,
Australia; brian@mso.anu.edu.au}

\altaffiltext{15}{Stockholm Observatory, AlbaNova, SE-106 91 Stockholm,
Sweden; jesper@astro.su.se}

\altaffiltext{15}{Hubble Fellow}

\begin{abstract}

We present photometric and spectroscopic observations of 23 high
redshift supernovae spanning a range of $z=0.34-1.03$, 9 of which are
unambiguously classified as Type Ia.  These supernovae were discovered
during the IfA Deep Survey, which began in September 2001 and observed
a total of 2.5 square degrees to a depth of approximately
$m\approx25-26$ in $RIZ$ over 9-17 visits, typically every 1-3 weeks
for nearly 5 months, with additional observations continuing until
April 2002.  We give a brief description of the survey motivations,
observational strategy, and reduction process.  This sample of 23
high-redshift supernovae includes 15 at $z\geq0.7$, doubling the
published number of objects at these redshifts, and indicates that the
evidence for acceleration of the universe is not due to a systematic
effect proportional to redshift.  In combination with the recent
compilation of Tonry et al. (2003), we calculate cosmological
parameter density contours which are consistent with the flat universe
indicated by the CMB (Spergel et al. 2003).  Adopting the constraint
that $\Omega_{total} = 1.0$, we obtain best-fit values of
($\Omega_{m}$,$\Omega_{\Lambda}$)=(0.33, 0.67) using 22 SNe from this
survey augmented by the literature compilation.  We show that using
the empty-beam model for gravitational lensing does not eliminate the
need for $\Omega_{\Lambda} > 0$.  Experience from this survey
indicates great potential for similar large-scale surveys while also
revealing the limitations of performing surveys for $z>1$ SNe from the
ground.

\end{abstract}


\section{Introduction}
\label{sec-intro}

\subsection{Searching for Cosmological SNe Ia--Past and Future}

It has now been over five years since the announcements by the
High-$z$ Supernova Search Team (Riess et al. 1998) and Supernova
Cosmology Project (Perlmutter et al. 1999) of evidence for the
acceleration of the expansion of the universe and the inferred
presence of a non-zero cosmological constant.  Since the implications
of this result are so profound for cosmology and our understanding of
fundamental physics, it has been the subject of intense scrutiny from
several different directions with attempts to test for further
confirmation or any sign of problems.

One goal of the immediate follow-up work was obtaining better
measurements of low and moderately-high redshift ($z<0.5$) Type Ia
supernovae (SNe Ia) to increase the confidence in their use as
standard candles for cosmological purposes.  Near-IR observations
(Riess et al. 2000) showed no evidence for extragalactic dust in a
single SN Ia at $z\approx0.5$, and spectra of a separate object at a
similar redshift (Coil et al. 2000) compared very closely with nearby
SNe Ia, showing no sign of spectroscopic evolution.  Sullivan et
al. (2003) demonstrated that host galaxy extinction is unlikely to
cause the observed dimming of high-redshift SNe, by comparing Hubble
diagrams as a function of galaxy morphology (see also Williams et
al. (2003) for a discussion of host galaxy-SNe correlations).
However, Leibundgut (2001) presented evidence that distant SNe Ia are
significantly bluer than the nearby sample, possibly indicating
photometric evolution that could bedevil analyses which assume that
color corrections can be made based on comparison to local SNe Ia.

There have also been continued attempts to discover supernovae at even
higher redshifts.  An extreme case is the serendipitous reimaging in
the Hubble Deep Field of SN 1997ff (Riess et al. 2001), which added
intriguing additional evidence for an earlier period of deceleration,
with the caveats that it is only a single object and potentially
gravitational lensed (Benit\'{e}z et al. 2002).  The sample size of
high-$z$ objects has been substantially added to by recent campaigns
described by Tonry et al. 2003 (8 SNe Ia between $0.3 < z < 1.2$), as
well as Knop et al. 2003 (11 SNe Ia between $0.36 < z < 0.86$).

The ability to discover large numbers of high-redshift supernovae with
reliability was made possible by the development of wide-field cameras
with large-format CCDs on large telescopes.  Observing time on these
instruments is extremely precious, and standard practice is to obtain
time for a template observation, followed some weeks later by a second
epoch from which to subtract the first epoch and thus detect
supernovae (see Schmidt et al. 1998).  The observations necessary to
obtain a complete photometric light curve of confirmed SNe Ia are then
made with other telescopes that can target individual objects, and on
which access to time is somewhat less competitive.  Spectroscopic
confirmation that a candidate is indeed a SN Ia requires significant
time on 8--10-m class telescopes, and the amount of such time that can
be obtained is often the limiting factor for supernova surveys.

The coming years will see a tremendous increase in the number of
astronomical surveys taking advantage of the ability of these
wide-field imaging cameras to cover large regions of sky.  In a new
twist, these surveys will observe large areas repeatedly in order to
explore the astronomical time domain in unprecedented ways.  This will
allow better understanding of a wide range of transient objects such
as asteroids, microlensing events, active galactic nuclei (AGN), and
supernovae, as well as potentially unveiling previously unknown time
variable phenomena.

This trend has already begun to a limited extent with such projects as
the Deep Lens Survey (Wittman et al. 2002) and Sloan Digital Sky
Survey (York et al. 2000), which in late 2002 began repeat coverage of
certain fields in order to search for variable objects (Miknaitis et
al. 2002).  Among other surveys underway is ESSENCE
(http://www.ctio.noao.edu/wproject, Smith et al. 2002), a five-year
program to discover hundreds of SNe Ia over a wide redshift range ($0.2 <
z < 0.7$) in order to measure the cosmological equation of state.  The
exploration of the wide-field, temporal-variability domain is
scheduled to culminate with truly massive undertakings such as the CFH
Legacy Survey (http://www.cfht.hawaii.edu/Science/CFHLS) and PanSTARRS
(http://poi.ifa.hawaii.edu, Kaiser et al. 2002).

The $Hubble\ Space\ Telescope$ ($HST$) has also recently entered the
fray with the Advanced Camera for Surveys (ACS; Ford et al. 1998),
giving it wide-field capability.  Several objects have been discovered
through observations of the Hubble Deep Field North (Blakeslee et
al. 2002), and in late 2002 a campaign was begun to discover
supernovae out to the redshift $z\approx1.7$ through strategic
placement of GOODS survey observations (Riess 2002), already yielding
numerous objects (Riess et al. 2003).  Finally, the extreme of
aspirations is the proposed Supernova Acceleration Probe (SNAP)
(Nugent 2001), a satellite mission specifically designed to discover
and monitor huge numbers of SNe Ia out to $z\approx1.7$.

\subsection{The IfA Deep Survey}

Beginning in September 2001, a collaboration of astronomers from the
Institute for Astronomy (IfA) at the University of Hawaii-Manoa
undertook the IfA Deep Survey, using wide-field imagers atop Mauna
Kea, Hawaii.  This project imaged 2.5 square degrees in multiple
colors ($RIZ$) roughly every 2--3 weeks for approximately 5 months,
with observations continuing until April 2002.  The major motivation
for separating the individual nights in this manner was to discover
and follow large numbers high-redshift supernovae.  The survey was
designed to accommodate investigations of a wide range of scientific
goals, including searches for substellar objects, galactic structure
studies, variable object searches (particularly supernovae), and
galaxy clustering studies.  Preliminary analysis of survey data has
already yielded at least one substellar object (Liu et al. 2002), and
scores of both high redshift supernova (Barris et al. 2001, 2002) and
brown dwarf candidates (Graham 2002; Mart\'\i n et al., in
preparation).

The novel feature of this campaign was the use of survey observations
to follow SNe Ia as well as find them.  No prior supernova campaign
has been performed in this manner.  At the beginning of any survey,
many supernovae will be discovered well past maximum light, which will
not be suitable for cosmological studies.  Similarly, supernovae which
are discovered before or at maximum light at the end of the survey
will not have sufficient follow-up observations to be useful.
However, all of the supernovae discovered in the middle of a
continuous survey will have observations on the rising portion of the
light curve as well as far into the decline, giving sufficient
coverage for light curve fitting and hence distance determination.

In this paper we describe the IfA Deep Survey and data reduction as
well as results from the supernova search.  In Section ~\ref{sec-obs}
we describe the survey observations.  In Section ~\ref{sec-red} we
give a brief description of the pipeline data reduction process, which
produces the final images to be used by all the collaborators.
Section ~\ref{sec-search2} describes the supernova search.  Sections
~\ref{sec-dist} and ~\ref{sec-anal} discuss the distance measurements
and cosmological analysis, and Section ~\ref{sec-conclusion} gives our
conclusions.

\section{Observations}
\label{sec-obs}
\subsection{Survey Science Observations}
\label{sec-science-obs}
For most of the scientific goals of the IfA Deep Survey, the primary
concern was overall survey depth.  The most important factors for the
supernova search component were sufficient depth on individual nights
to detect high-redshift supernovae and separation of the nights so as
to allow for continual detection and follow-up throughout the duration
of the survey.  The primary instruments used were Suprime-Cam
(Miyazaki et al. 1998) on the Subaru 8.2-m telescope, and the 12K
camera (Cuillandre et al. 1999) on the Canada-France-Hawaii 3.6-m
telescope (CFHT).

The survey strategy was designed to provide for the discovery and
follow-up of 10--25 SNe Ia with $0.9<z<1.2$ in order to distinguish
whether the evidence from SNe Ia at lower redshift for an accelerating
universe could actually be due to a systematic effect proportional to
redshift rather than an indication of a non-zero cosmological
constant.  The rates of high-redshift supernovae are still quite
uncertain (see Pain et al. 1996, 2002 and Tonry et al. 2003), but the
rate of SNe Ia in our desired redshift range is approximately 2--5 per
sq. deg. per month, depending on when the individual images are taken.
The peak brightness of a SN Ia at $z=1.2$ is about $I=24.3$ and
$Z=23.6$, so each survey night was designed to provide a
signal-to-noise ratio (S/N) of 11 in $I$ band and S/N = 7 in $Z$ band
at these magnitudes, assuming 0.75\arcsec\ seeing.  This is a bit less
S/N than ideal for $Z$ band, but some observations were taken with
much better seeing which went deeper (and conversely, some were taken
with worse seeing and therefore went shallower).  In addition, the $R$
band images, designed for S/N = 10 per observation, can be used to
help distinguish SNe Ia from SNe II, based on the redder color of
high-redshift SNe Ia (see Section ~\ref{sec-likely} below).

Fields and observational cadence were chosen to meet the necessary
requirements of all the scientific programs.  Primary considerations
included a spread in right ascension to allow for continuous
observation throughout a single night over several fall and winter
months from Mauna Kea; low galactic extinction; and whether previous
observations, in the same or other wavelength regions, could be used
to augment the scientific objectives.  Five 0.5-square degree fields
were chosen for the survey.  Central coordinates of the selected
fields are given in Table ~\ref{table:coords}.  The field at $02^{h}27^{m}$
(Field 0230) was previously used for high-redshift surveys by
the High-$z$ Supernova Search Team (see Tonry et al. 2003).  Field
0848 was chosen to overlap with previous radio observations, and Field
1052 (``Lockman Hole,'' see Lockman, Jahoda, \& McCammon 1986) was
chosen for a wide range of prior multi-wavelength studies (x-ray:
Hasinger et al. 1993; radio: de Ruiter et al. 1997, and Ciliegi et
al. 2003; IR: Taniguchi, Kawara, \& Matsuhara 1999, and Fadda et
al. 2002).

The 12K camera consists of twelve 2048x4096 pixel CCDs, with a field
of view of $45'$ x $30'$ (0.375 sq. degrees) and a pixel scale of
$0.206''$/pixel.  Survey fields were observed with a Mould $I$ filter,
with a central wavelength of 8223 \AA\ and a width of 2164 \AA.

\begin{figure}
\plotone{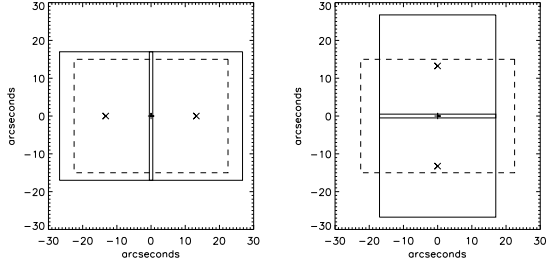}
\caption{Configuration of the Suprime-Cam and CFHT+12K fields of view during
the IfA Deep Survey.  Solid lines depict the two Suprime-Cam fields of
view, dashed lines show the coverage of the single 12K field of view.
Shown on the left is the layout for Fields 0230, 0438, 0749, and 0848.
Shown on the right is the layout for Field 1052.}
\label{fields}
\end{figure}

The Suprime-Cam instrument, consisting of a mosaic of ten 2048x4096
pixel CCDs, covers a $34'$ x $27'$ field of view (0.255 sq. degrees),
and has an image scale of $0.20''$/pixel.  The survey fields were
observed with Suprime-Cam with Cousins $R$ and $I$ and Subaru $Z$
filters.  The $Z$ filter at Subaru has an effective wavelength of 9195
\AA\ and FWHM of 1410 \AA\ (Fukugita et al. 1996).  Each of the five
fields was covered by two overlapping Suprime-Cam fields-of-view
(FOVs).  The central coordinates for the pair of Suprime-Cam pointings
for each field are given in Table ~\ref{table:coords}.  For four of
the fields (all except for Field 1052), the two Suprime-Cam
fields-of-view were rotated by 90 degrees relative to the 12K FOV.
Thus the coverage with two Suprime-Cam fields-of-view was
approximately $34'$ x $54'$ ($\approx$0.5 square degrees), compared to
$30'$ x $45'$ (accounting for the orientation) with the 12K.  For
field 1052 the appropriate comparison is $34'$ x $54'$ with
Suprime-Cam and $45'$ x $30'$ with 12K, so that in this configuration
there were regions on the edge of the field which were imaged with the
12K but not with Suprime-Cam, and vice versa, while for the other
fields the 12K FOV is completely covered by the Suprime-Cam footprint
(see Figure ~\ref{fields} for an illustration).

The bulk of the survey took place over 8 full nights and 5 half nights
with Suprime-Cam on Subaru from October 2001 through April 2002.
Target fields were also observed between 1 and 5 times with
queue-scheduled observations with the 12K at CFHT.  Not all nights
were photometric, and exposure times occasionally varied depending on
conditions.  A summary of observations, including date, exposure time,
and mean seeing value, is given in Tables ~\ref{table:cfhobservations}
(CFHT 12K) and ~\ref{table:subaruobservations} (Suprime-Cam).

\begin{figure}     
\plotone{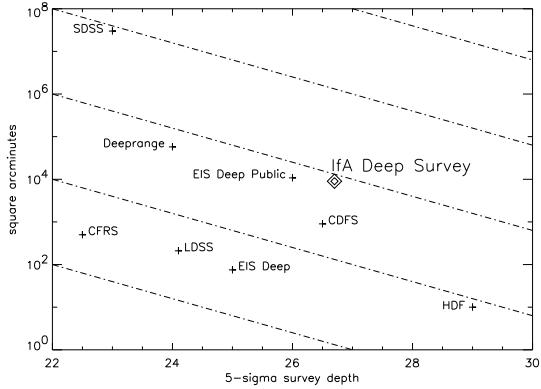}
\caption{A plot of depth (approximate AB magnitudes) versus size for
several recent and historical surveys.  Diagonal lines show contours
of constant volume.  The IfA Deep Survey covers a unique region of
this parameter space.  SDSS (Sloan Digital Sky Survey, www.sdss.org,
York et al. 2000).  Deeprange (Postman et al. 1998).  EIS (ESO Imaging
Survey) Deep, Deep Public (www.eso.org/eis).  CFRS (Canada-France
Redshift Survey, Lilly et al. 1995).  LDSS (Low Dispersion Survey
Spectrograph, Glazebrook et al. 1995).  CDFS (GOODS/ESO Chandra Deep
Field South, www.eso.org/eis).  HDF (Hubble Deep Field, Williams et
al. 1996).
\label{limits}}
\end{figure}

Exposure times were chosen to achieve approximately the same survey
depth in the three filters, though the required exposure times for $Z$
were impractically long, and so this filter did not go as deep as $R$
and $I$.  The long readout time for Suprime-Cam was also a major
factor.  For a typical night, 5-$\sigma$ point source sensitivities
were roughly 25.8 in $R$, 25.2 in $I$, and 24.2 in $Z$.  When the
entire survey is combined, the 5-$\sigma$ point source depth of the
summed images is approximately 27.3 in $R$, 26.7 in $I$, and 25.6 in
$Z$, varying from field to field due to differences in integration
time as well as seeing conditions on nights when a particular field
might be more heavily represented.  See Figure ~\ref{limits} for a
comparison of the depth and area coverage of the IfA Deep Survey with
several other recent surveys, illustrating where this survey lies in
area vs. sensitivity parameter space.

\subsection{Astrometric Observations}
\label{sec-astro-obs}

In addition to the survey science observations, we also obtained
shorter exposures in order to construct astrometric catalogs of the
target fields.  These observations, taken with the CFHT 12K camera,
were 120 s in all bands (Mould $R$, Mould $I$, $Z$).  One advantage of
these shorter images is that they create a photometric overlap with an
external reference, the USNO-A catalog (Monet 1998), since the much
deeper science images discussed in the previous section do not contain
any non-saturated stars from this catalog.  The images of each field
were taken with half-CCD offsets to determine astrometry over the
entire survey area.

Object detection software was run on the astrometric images (findpeaks
from the imcat package, see http://www.ifa.hawaii.edu/$\sim$kaiser) to
construct a catalog of objects to register with the USNO-A catalog.
Solutions for image mapping parameters were then obtained using a
process developed for weak lensing studies (see Kaiser (2000) for a
detailed mathematical description of the method, and Kaiser et
al. (1999) for a more practical summary).  Stars brighter than
$m\approx21$ are identified in the astrometric images as well as the
USNO-A catalog, and a large matrix equation mapping CCD coordinates to
sky tangent plane coordinates is fit by a cubic polynomial.  We then
iterate until the process converges, rejecting outlying points after
each step.  Great care was needed so as not to mistakenly reject large
groups of stars in a given region which may show a systematic offset
due to a single problematic star.

After achieving an acceptable solution (typically requiring 4--6
iterations), more sensitive object detection was run on the
astrometric images to augment the catalog with stars faint enough to
overlap with the much deeper science images.  Going deeper means that
many objects in the reference catalog will be faint galaxies rather
than stars, but since they are stellar in appearance they are still
suitable for astrometry.

The end product of this process was a catalog of stars with extremely
accurate relative astrometry (to a fraction of a pixel, i.e. better
than 0.1 arcseconds), extending to very faint magnitudes ($m\approx22$
for $R,I$ and $m\approx21$ for $Z$).  The catalogs range in size from
about 4000 stars for Fields 0230, 0848, and 1052; to 7000 for Field
0438; to more than 9500 for Field 0749, which is at a lower galactic
latitude than the other fields.

\subsection{Photometric Observations}
\label{sec-photo-obs}

Photometric observations were obtained at the CTIO 1.5-m and UH 2.2-m
telescopes.  Landolt standards (Landolt 1992) and spectrophotometric
standards, which have Landolt magnitudes as well as synthetic $Z$
magnitudes, were observed in $BVRIZ$ (Johnson $BV$, Cousins $RI$, and
$Z$ as described in Tonry et al. 2003) to set the magnitude scale of
the survey target fields.

Flux measurements for the standard stars were calculated using $14''$
diameter aperture magnitudes of isolated local standards in each
field.  For each night and filter an atmospheric extinction
coefficient and a color term was calculated.  This fit to airmass and
color typically showed a scatter of 0.02 mags, due to the usual
difficulties of data reduction: sky errors, flatfielding
imperfections, atmospheric transparency variations, CCD
non-linearity, shutter timing errors, and variations due to PSF or
scattered light.

Our $Z$ band observations were calibrated by observing a series of
Landolt stars, whose magnitudes were derived by integrating their
spectrophotometry (Suntzeff, in preparation) with our bandpass
defined by the CTIO natural system.  This system is defined to have
$(V-Z)=0$ for Vega.

The astrometric catalogs described above have accurate relative
photometry, and the CTIO 1.5-m and UH 2.2-m observations allowed us to
put them on an accurate absolute scale.  Calibration of the 0230 field
was described by Tonry et al. (2003), wherein cross-checks with the
Sloan Digital Sky Survey (SDSS, Stoughton et al. 2002) indicate an
accuracy as good as the SDSS zero-point uncertainty of 0.04
magnitudes.

\section{Pipeline Reductions}
\label{sec-red}

Survey science images were initially reduced using a reduction
pipeline created specifically for this purpose at the IfA.  The
majority of the pipeline consisted of custom scripts using the Vista
image display and manipulation software.  The search for high-redshift
supernovae was a time-critical mandate for the reductions, so the
pipeline was designed to be as efficient as possible.  Initial
processing took place as soon as the data were received via {\scshape
ftp} from the telescope.  A second processing, free of time pressures
and with slight modifications, was performed after the conclusion of
the survey.

The images were first bias subtracted using a median value from pixels
in the overscan region of each chip.  For the Subaru data, a median
superflat image was constructed from all the images for a given chip
and filter from the entire night.  Images were flattened by dividing
by this superflat.  For the CFHT 12K images, dome flats were used to
flatten the data, and a fringe frame was constructed from the entire
night.  The images were flattened by the dome flat image, and the
fringe frame subtracted.  The Suprime-Cam observations had small
enough fringing (less than a few percent) that division of the fringe
light instead of subtraction did not significantly affect photometry.
Bad pixels and other chip defects were removed using a mask created at
the start of the campaign, which proved to be sufficient for the
entire survey.  Any remaining tilt in the sky was subtracted, and the
sky normalized to a value of 1000 for ease of software compatibility.

The flattened images were then mapped to the astrometric catalogs of
the survey fields through detecting (via SExtractor) and matching
stars (using a similar process as described for the astrometric
observations, though now the custom-made astrometric catalogs were the
reference, rather than the USNO-A catalog).  Using these astrometric
solutions, the images were warped onto a predefined coordinate system
on a sky tangent plane.  The final post-warp images used a pixel scale
of $0.20''$/pixel, which is similar to both Suprime-Cam and CFHT 12K.
The warping process conserves flux and uses a Jacobian to restore
photometric accuracy lost by flatfielding.  Since dividing by the
flatfield in the pre-warp stage equalizes surface brightness
regardless of geometrical distortion of pixel area, and our warping
conserves flux rather than surface brightness, the Jacobian is
necessary for accurate photometry.

At this point cosmic ray (CR) rejection was done and the images
combined.  Performing CR rejection at this stage has the drawback of
decreased sensitivity due to smearing of the cosmic rays during the
warping, and thus many were not removed.  For the goal of discovering
supernovae, this was not an insurmountable problem, as individual
dithers contributing to the final image of a candidate could be
inspected.  In the second processing, performed after the conclusion
of the survey, the cosmic ray rejection was performed before, rather
than after, the warping stage.  Images were registered by adaptively
finding subimages for which integer pixel shifts gave sufficiently
close registration, and cosmic ray rejection performed after the
processing steps of flattening, pixel masking, etc.  This resulted in
a significant improvement in the removal of cosmic rays.  After the CR
rejection, the warping was performed with cosmic ray masked pixels
treated the same as other bad pixels.

\section{Supernova Search}
\label{sec-search2}

\subsection{Supernova Discovery}
\label{sec-discover}

In order to perform the supernova search, previous $I$ observations
were subtracted from each new set of images to detect photometrically
variable objects.  Because our first set of observations, in September
2001, were taken with the CFHT 12K, subtractions from our first
Subaru+SuprimeCam nights in October could only be done over the area
in common with both telescope fields-of-view (the central 0.375
square-degrees of the 0.51 square-degrees covered by SuprimeCam).  In
subsequent months we possessed a complete set of templates which
allowed for searching of the entire survey area.

Subtractions were performed largely in the same manner as for previous
supernova searches (see Schmidt et al. 1998, Tonry et al. 2003).  New
software was written (Becker et al., in preparation), slightly
modifying the algorithms of Alard (2000) and Alard \& Lupton (1998)
based on insights gleaned from searches for microlensing as well as
past supernova campaigns.  Pairs of observations to be subtracted were
compared to determine the worse seeing night, with the better night
then convolved to match PSFs.  Flux levels were matched, and the
images subtracted.

Two independent automatic search algorithms were run on the subtracted
images to detect variable sources.  The images were divided into
regions 7 arcminutes on a side for ease of searching, and in a typical
such region several dozen variable sources were detected.
Approximately half of these were false positives such as diffraction
spikes around saturated stars, or filter mismatches between CFHT and
Subaru.  A large number were variable stars and AGN, which are
typically distinguished from supernovae by being a small fluctuation
centered on a constant point source.  All subtractions were searched
visually by at least one person, aided by the automatic algorithms.
In addition to the many false positives, the automatic detection
programs missed some faint supernovae, but in general seemed to do a
respectable job in finding objects when compared to well trained
searchers.  After the initial search was completed, candidate
supernovae were inspected by an additional observer.  The constituent
images (typically a single $I$ band observation was comprised of 3
separate dither positions) were also examined to weed out cosmic rays
and moving objects.

Ideally, the fact that we had observations in three filters would have
been extremely useful in distinguishing between SNe Ia and SNe II
based on the difference in color at early times (at high redshifts,
SNe II are much bluer than SNe Ia in $RIZ$).  In practice, however,
reducing and searching the $I$-band data alone consumed all available
resources.  Upon further inspection of the IfA Deep Survey
observations we should be able to provide an excellent assessment of
whether theoretical photometric discriminatory tests (see Poznanski et
al. 2003) are actually useful for real-world surveys.  We have already
begun to use them for the purposes of augmenting our sample, as
described in Section ~\ref{sec-likely} below.

Our continuous search strategy did enable us to attempt to
differentiate between the types of supernovae based on the shape of
their early light curve.  Type II SNe have extremely heterogeneous
light curves (see Leibundgut \& Suntzeff 2003, and references
therein).  One major subclass, the SNe II-P, exhibit a very short
rise to maximum light, followed by a plateau in brightness.  After a
detection we were usually able to compare with an observation made a
week or two previously.  Quite often the supernova was present in this
prior epoch but faint enough that it was not detected by our
searchers, indicating a gradual rise, and hence a likely SN Ia rather
than a SN II.  Alternatively, if the object was not present at all in
this previous epoch, it was more probably a SN II or a SN Ia at a much
lower redshift.

After this rigorous inspection process candidates were prioritized and
forwarded to the spectroscopic observers, who were usually already
waiting at the telescope.

\subsection{Spectroscopic Observations}
\label{sec-spec}

Though the continuous nature of the IfA Deep Survey naturally provided
for photometric follow-up of supernova candidates, guaranteeing
successful spectroscopic follow-up was still problematic.  The only
possibility for supernova surveys is to rely on time allocation
committees to schedule observing nights a few days after the
photometric search observations and hope that supernova candidates for
spectroscopy will be ready in time.  Complications can arise from many
uncontrollable issues---the photometry night may have poor conditions,
the searchers may encounter hardware and/or software problems, the
timing between photometric and spectroscopic nights may end up being
too short, among other pitfalls.  Despite the difficulty involved,
measurement of redshift and identification as a SN Ia is necessary to
be able to use a supernova for cosmological purposes.

We arranged spectroscopic resources on many telescopes over the course
of the IfA Deep Survey---primarily the Keck I and II telescopes, with
the VLT also playing a major role, though it was unable to reach our
northernmost fields.  Additional spectroscopic observations were made
with the Magellan telescope, though it was not used for identification
of any of the high-$z$ SNe Ia discussed here.

We observed a total of 63 objects spectroscopically during 6 separate
observing runs (4 with Keck, in addition to one each with the VLT and
Magellan) spaced throughout the duration of the survey.  We were also
able to observe a small number of supernova host galaxies in late 2002
during the first year of the ESSENCE project (Smith et al. 2002).

\subsubsection{Keck-II + ESI} 
\label{sec-esiobs}
                       
We obtained spectra of supernova candidates with ESI (Sheinis et
al. 2002) on the Keck-II telescope during two separate observing runs
in October and November of 2001.  On all nights we used a $1.0''$ slit
for the observations, with a spectral resolution of $\sim$1.4 \AA .
The slit was oriented along the parallactic angle, or to include
either the nucleus of the host galaxy or a nearby bright star.
Between integrations the target was moved along the slit to reduce the
effects of fringing and increase confidence in identifying the SN
spectrum.  On October 21-22, the seeing was $0.6''-0.8''$.  For
November 16-18, the seeing was $0.5-0.6''$ on the first night and
$0.7-0.8''$ on the final two nights.  The standard stars BD+174708 and
BD+284211 (Oke 1990) were used for flux calibration.  Standard CCD
processing and optimal spectral extraction were done with IRAF, with
our own IDL routines used to calibrate the wavelengths and fluxes of
spectra and to correct for telluric absorption bands.

We also obtained spectra with ESI on November 6, 2002, during ESSENCE
project observations in order to measure additional redshifts of host
galaxies.  Seeing was $0.8''-0.9''$, and observational setup and
reduction was the same as described for the 2001 nights.

\subsubsection{Keck-I + LRIS}
\label{sec-lrisobs}

Additional spectra of SN candidates were obtained with LRIS (Oke
et al. 1995) on the Keck-I telescope during two separate observing
runs in December 2001 and January 2002.  On December 22, the seeing
was approximately $1''$, with high cirrus.  On January 16-17, the
seeing was $1.1-1.8''$ on the first night and $2-3''$ on the second
night.  We used a $1.0''$ slit for the observations.  The spectral
resolution is $\sim$6.2 \AA\ for the red end (6500-10000 \AA ), and
9.2 \AA\ for the blue (3300-6900 \AA ).  Between integrations we moved
the target along the slit for the reasons described above.  In
December standard star BD+174708 (Oke \& Gunn 1983) was used for flux
calibration, while in January standard stars BD+174708, BD+262606, and
BD+284211 (Oke 1990) were used.

As with the ESI observations, we used IRAF for standard CCD processing
and spectral extraction, and IDL for calibrating the wavelengths and
fluxes of spectra and removing atmospheric absorption bands.  We
used the 300/5000 grism on the blue side over the range 3300-6900 \AA
, which was matched to the D680 dichroic.  On the red side, we used
the 400/8500 grating over the wavelength range 6500-10000 \AA .
Typically the two sides were tied together over the range 6500-6600
\AA .  There were any number of anomalies in the overlap range, which
were minimized as much as possible, but some complications were
unavoidable.  Specific problems encountered included (1) second-order
light on the blue side; (2) wavelength-dependent time variations in the
dichroic; and (3) reflections from bright stars landing on targets on
the blue side.  We were able to work around the second-order blue
light in our standard stars and ignore it in our objects (which are
red).  We also corrected for the variations in dichroic transmission
to the red side, but not in the reflectance to the blue side.  There
is an increase in noise and possible spurious features exist near the
overlap region.

\subsubsection{VLT + FORS} 
\label{sec-vltobs}

We also obtained spectra of SN candidates with FORS1 on the VLT during
an observing run on December 13 and 14 of 2001. On December 13, the
seeing varied from $0.7''$ to $1.8''$. On December 14, the seeing
changed rapidly from $\approx0.7''$ to $1.3''$ until heavy clouds
moved in mid-way through the night, so that most of the subsequent
data did not have enough signal for analysis.  We used a $1.0''$ slit
for all observations, with a 300 line mm$^{-1}$ grating.  The slit was
oriented to include the nucleus of the host galaxy. FORS1 is mounted
behind an atmospheric dispersion compensator and we do not expect any
wavelength dependent effect due to the setting of the slit at an angle
different from the parallactic angle.  Between integrations the target
was moved along the slit to reduce the effects of fringing and for
ease of identifying the SN spectrum.  During the night of December 13
we observed EG21 and LTT 3218 as spectroscopic standards (Hamuy et
al. 1992, 1994) at the beginning and end of the night, respectively.

The spectra were first corrected by bias subtraction and division by a
dome flat. The wavelength scale was established through calibration
observations of a HeCuAr lamp observed during the afternoons before
and after the observations. The wavelength solution was then verified
against night sky lines. The extraction of the spectra was carefully
done to control the contamination by the underlying host galaxies.
The spectra were then flux-calibrated using the sensitivity functions
derived from the standard stars.

\subsection{Supernova Classification}
\label{sec-likely}
Type Ia supernovae are defined by the presence of Si {\scshape ii}
$\lambda6355$ and Ca {\scshape ii} H+K $\lambda\lambda3934, 3968$, and
typically show numerous broad undulations in their spectrum (see
Filippenko 1997).  For the redshifts targeted during this survey, the
Si {\scshape ii} line was not observable with visible spectroscopy,
though we often could identify the Ca features, 4000 \AA\ break, and
other SN Ia features.  Among the features that may be used to uniquely
identify a SN Ia when Si {\scshape ii} is not observed is the
double-bump feature near 4000 \AA\ created by the combination of Fe
{\scshape ii} $\lambda4555$ and Mg {\scshape ii} $\lambda4481$.
Typically the redshift was measured through host galaxy emission, most
often the {\scshape [O ii]} doublet at $\lambda\lambda3726, 3729$.
This doublet is often unresolved by many instruments, and so appears
as a single prominent feature, which in many cases could be
interpreted as {\scshape [O ii]} at high redshift, or H$\alpha$ at low
redshift.  One means of distinguishing between these choices is the
brightness of the host galaxy.  Another is that H$\alpha$ is likely to
be accompanied by such lines as {\scshape [N ii]}, {\scshape [S ii]},
or {\scshape [O iii]}, so the lack of these other features indicates
that {\scshape [O ii]} is more likely.  The ability to resolve this
doublet will remove all doubt, and this was one advantage of using
ESI, with its much greater resolving power compared to LRIS for
spectroscopic observations (note the values for each instrument given
above in Sections ~\ref{sec-esiobs} and ~\ref{sec-lrisobs}).

Supernovae were identified by matching spectral features with those of
SNe Ia through a program called SNID (Tonry, in preparation), which
cross-correlates an observed spectrum against a set of template
spectra to determine the supernova type, redshift, and age.  SNID uses
a set of 171 template spectra which span a large range of SN Ia
properties (a small number of templates for other types of supernovae
are also used, though the diversity of SNe Ib/c and SNe II makes their
usefulness more limited than for the relatively homogeneous SNe Ia).
The best matches are reported in terms of a correlation value $r$.
Comparisons at very different redshifts may require trimming the
spectrum to varying degrees, and so SNID weights the correlation value
by the amount of overlap between the spectrum and template
(abbreviated $lap$), producing a parameter $r*lap$ which is used to
determine which template matches are of highest significance.

\begin{figure}
\epsscale{2.2}
\plotone{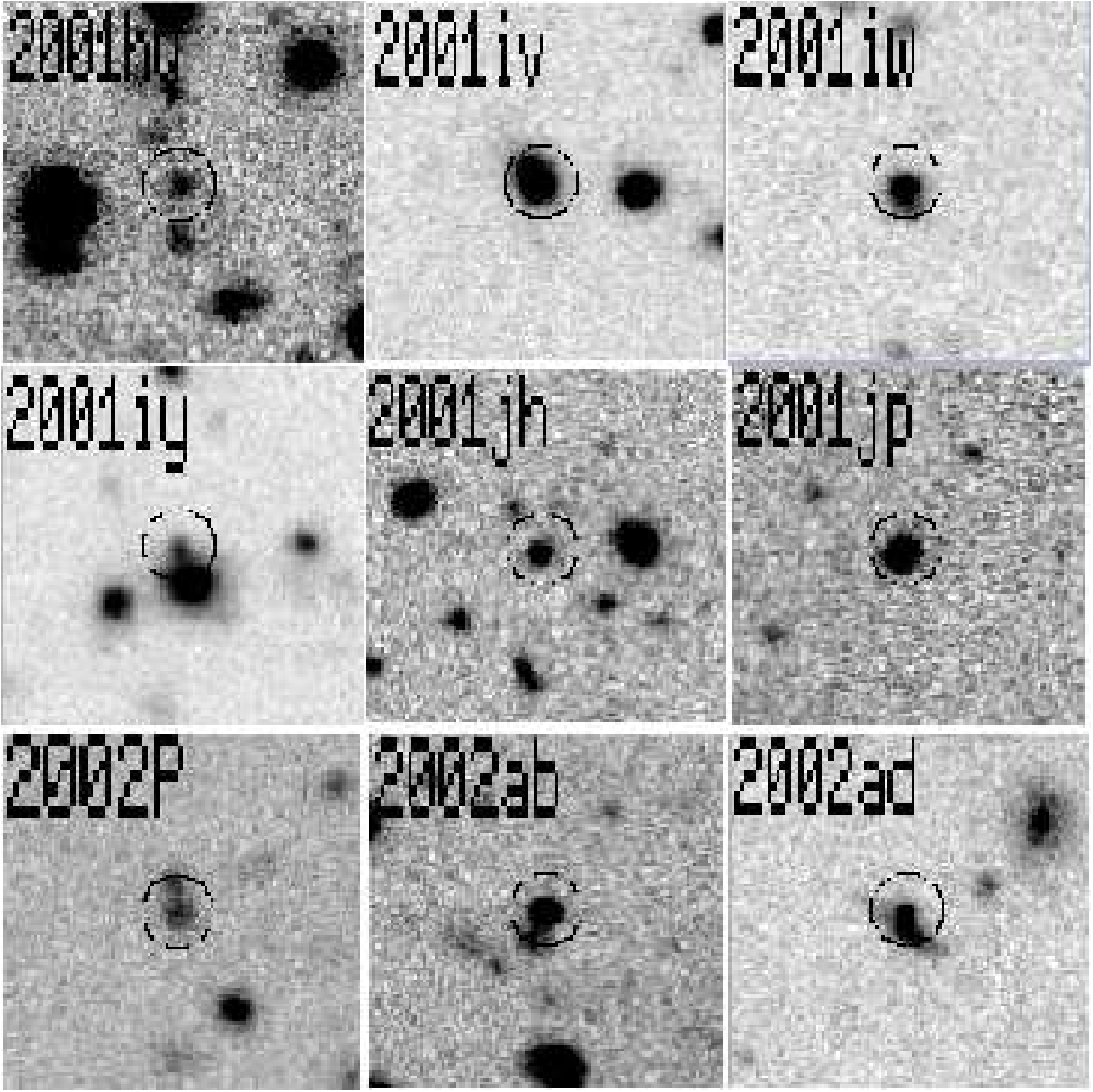}
\caption{$I$-band Subaru images centered on the location of each of
the 9 IfA Deep Survey SNID-confirmed SNe Ia (indicated with a circle),
taken as close to peak brightness as possible.  Images are $20''$ on a
side.  North is up and East to the left.}
\label{iapics}
\end{figure}

\begin{figure}
\epsscale{2.2}
\plotone{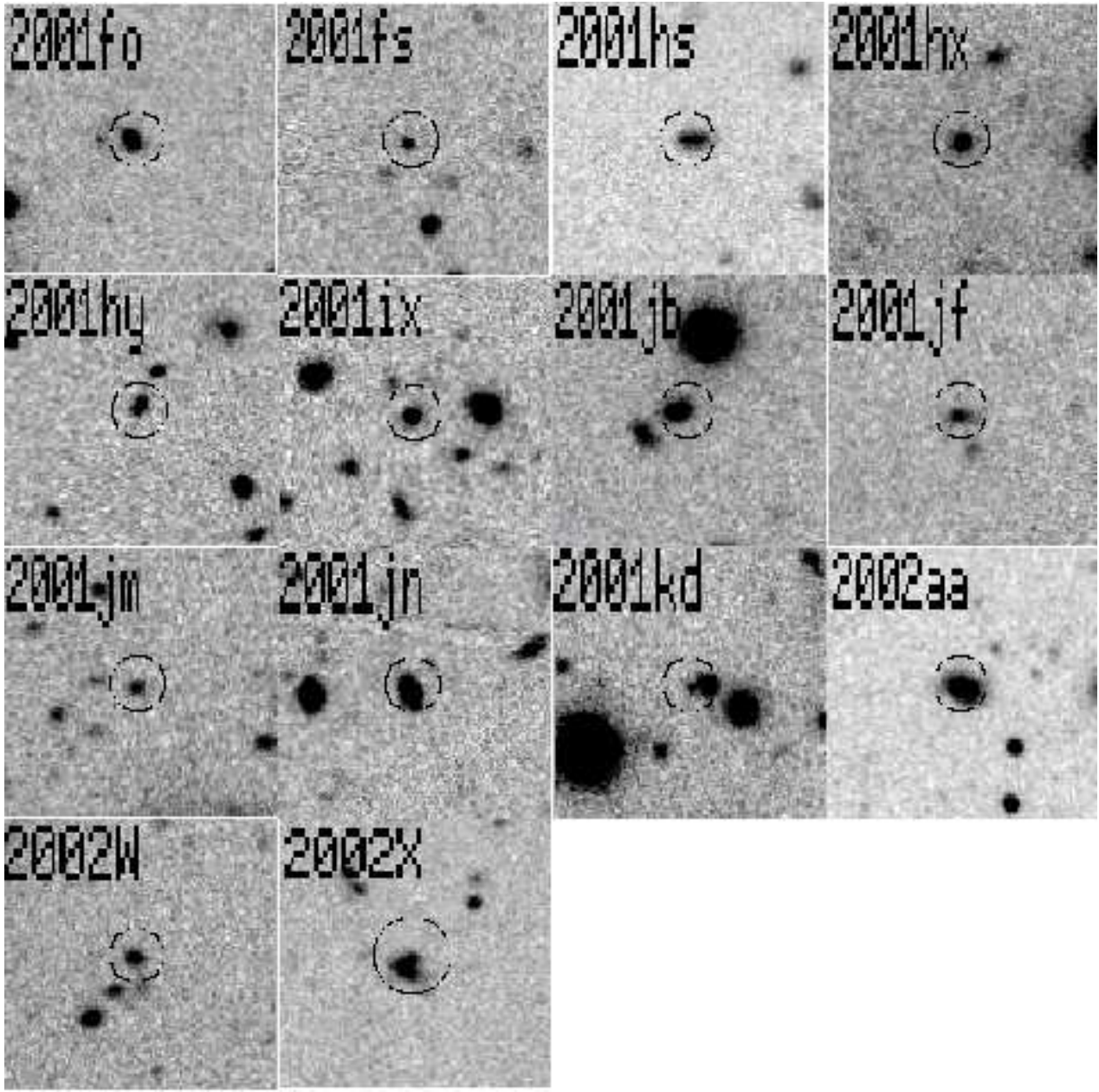}
\caption{$I$-band Subaru images centered on the location of each of
the 14 additional SNe Ia (indicated with a circle), taken as close to
peak brightness as possible.  Images are $20''$ on a side. North is up
and East to the left.}
\label{otherpics}
\end{figure}

\begin{figure}
\epsscale{2.2}
\plotone{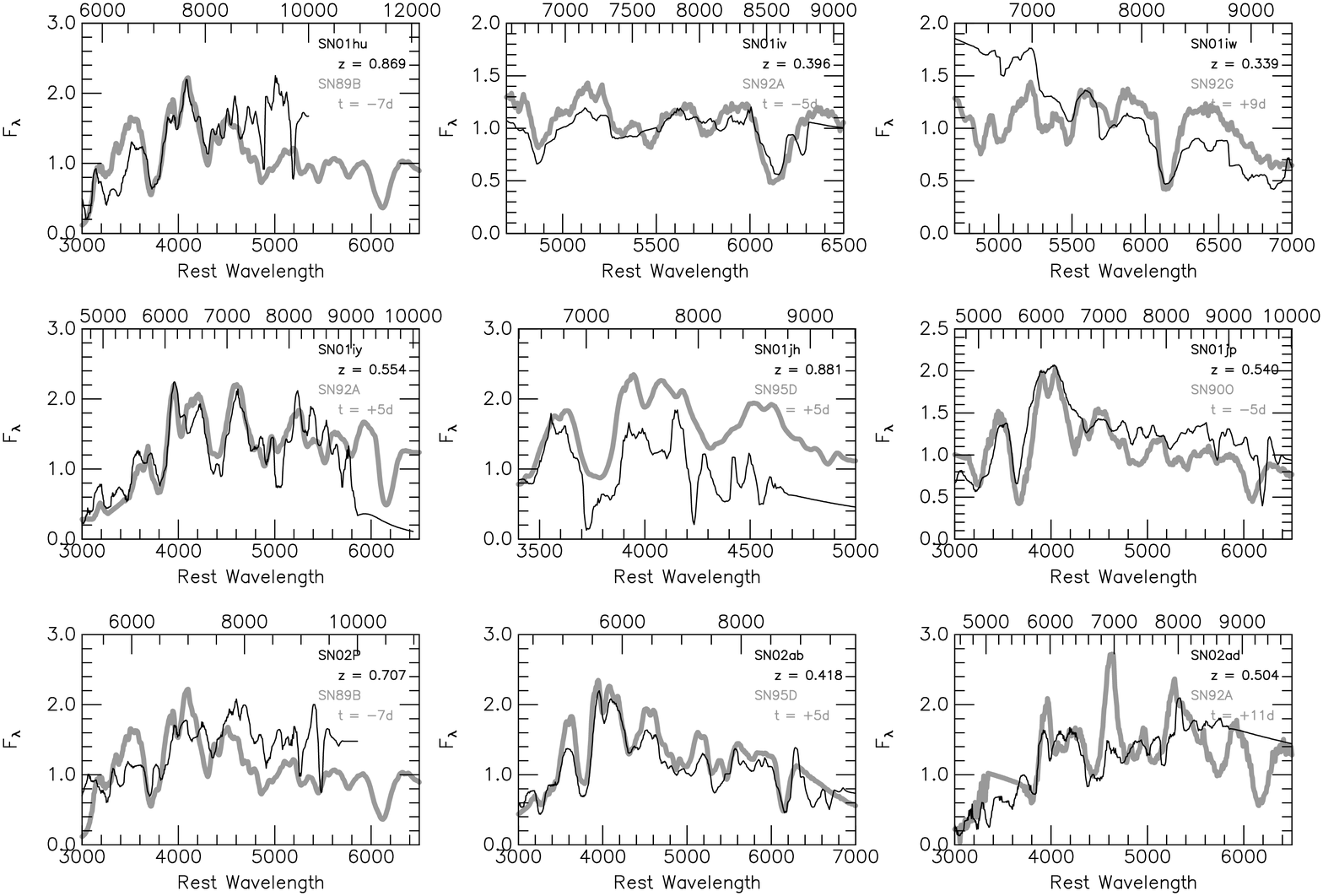}
\caption{Spectra of the 9 SNe with spectral matches to local SNe Ia as
determined by SNID.  The spectra have been smoothed by taking a
weighted median of FWHM 80 \AA.  Both spectra and template are shown
as F$_{\lambda}$. Observed wavelength is indicated along top of
graphs, with all wavelengths given in angstroms. }
\label{ia-spec}
\end{figure}

Table ~\ref{table:sninfo} contains positions and galactic extinction
values for 23 supernova candidates which we believe are truly SNe Ia,
with spectroscopic observations described in Table
~\ref{table:spectralist} and images shown in Figures ~\ref{iapics} and
~\ref{otherpics}.  Tables ~\ref{table:spectraid} and
~\ref{table:snidinfo} give information on the redshift determination
and SNID analyses, respectively.  Nine objects are identified as
unambiguous SNe Ia, meaning that they produce a strongly significant
best-fit SNID correlation value at the same redshift as indicated by
their host galaxy emission lines.  The exception is SN 2002ad, for
which the best SNID value is at $z\approx0.77$, while the second most
likely value is $z=0.514$, which agrees with the host galaxy emission.
However, the match at $z=0.765$ is with a template from just past
maximum light, whereas the $z=0.514$ match is at nearly two weeks past
max, agreeing with the light curve.  We therefore include SN 2002ad as
an unambiguous SNe Ia at the host galaxy redshift.  Spectral matches
for these 9 SNe Ia as determined by SNID are shown in Figure
~\ref{ia-spec}.  Note the presence of Si {\scshape ii} $\lambda6355$
in SNe 2001iv and 2001iw, and evidence of the double peak at
$\sim$4000 \AA\ in the other seven SNe, indicating that these are
indeed SNe Ia.

\begin{figure}
\epsscale{2.2}
\plotone{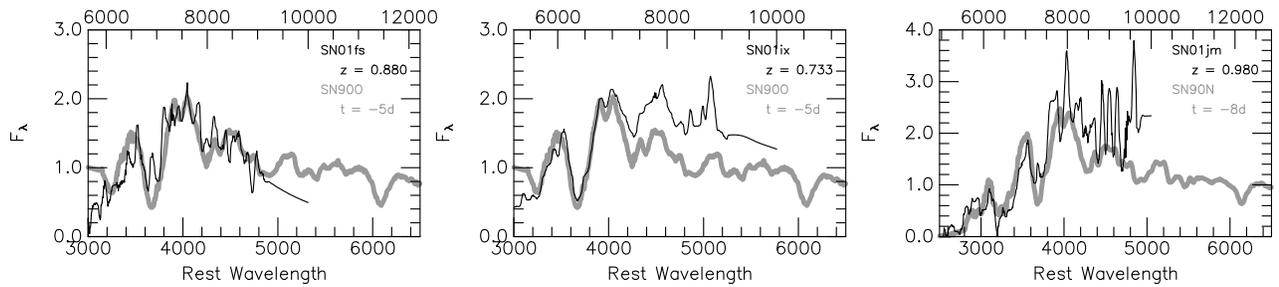}
\caption{Spectra of the 3 SNe without a host-galaxy emission redshift
but with spectral matches to local SNe Ia as determined by SNID.  The
spectra have been smoothed by taking a weighted median of FWHM 80 \AA.
Both spectra and template are shown as F$_{\lambda}$.  Observed
wavelength is indicated along top of graphs, with all wavelengths
given in angstroms.}
\label{three-spec}
\end{figure}

The remaining 14 objects are divided into two groups---11 with host
galaxy redshifts which do not correspond to a strong SNID match, and 3
for which there is a significant SNID correlation with a SN Ia
template but no host galaxy emission with which to compare.  Of the
group of 11, four (SNe 2001fo, 2001hs, 2002W, 2002X) have peak values
of $r*lap>3.0$, and so based purely on SNID correlation values appear
to be as strong SN Ia candidates as SN 2002ad.  The spectral matches
are not convincing, however, and we are unwilling to accept these fits
as significant (contamination of a spectrum by galaxy light often
makes it impossible to clearly discern SN features, so they are still
possible SNe Ia).  One of the 11 objects (SN 2001jn) was observed in
November 2002, long after the supernova had faded from view, so that a
SNID analysis is not possible.  The final group of 3 SNe (SNe 2001fs,
2001ix, 2001jm) do not possess visible host galaxy emission, but do
produce convincing matches with template SNe Ia via SNID, as shown in
Figure ~\ref{three-spec}.  However, SNID often produces a series of
correlation peaks at discrete redshifts, due to successively matching
different features in the spectrum of a SN Ia with alternate template
features at different redshifts.  Because of this ambiguity, we are
unwilling to state purely on a spectroscopic basis that these objects
are SNe Ia at the quoted redshift.  Inspired by our substantial
high-quality photometric information, we have pursued further
investigations to determine whether they may be added to our sample
with confidence that they are indeed SNe Ia.

A first test of these additional objects is to compare their
photometric observations with what one would expect from various types
of supernovae at the measured redshift.  SNe Ia have been subjected to
sufficient scrutiny in recent years that there is a good understanding
of their general properties and a large number of well-observed
objects.  We have used two examples to demonstrate the breadth of
parameter space that SNe Ia can be expected to cover---SN 1995D (Riess
et al. 1999b) is a bright SNe Ia, with $\Delta=-0.42$ as measured by
Riess et al. (1998), and $\Delta=-0.44$ from the MLCS method from this
paper (see Section ~\ref{sec-mlcs}); SN 1999by is one the most
subluminous SNe Ia ever observed (Garnavich et al. 2003).  At redshift
$z>0.8$, objects such as SN 1999by will be too faint for detection at
the $m\approx24.3$ $I$-band sensitivity of our survey, but it can
still serve as an illustrative limiting case for faint SNe Ia.

SNe II are more difficult because they are a less uniform population
than SNe Ia (see Leibundgut \& Suntzeff 2003).  We have selected
single spectroscopic observations of SN 1998S (Lentz et al. 2001;
Leonard et al. 2000) and SN 1999em (Baron et al. 2000; Leonard et
al. 2002) to serve as our templates.  SN 1998S is a SN IIn which was
observed at maximum brightness, and SN 1999em is a SN II-P which was
also observed near maximum.  These two objects can only begin to
describe the vast diversity of SNe II, and were chosen primarily for
their extensive spectral coverage into the UV.

\begin{figure}
\epsscale{2.2}
\plotone{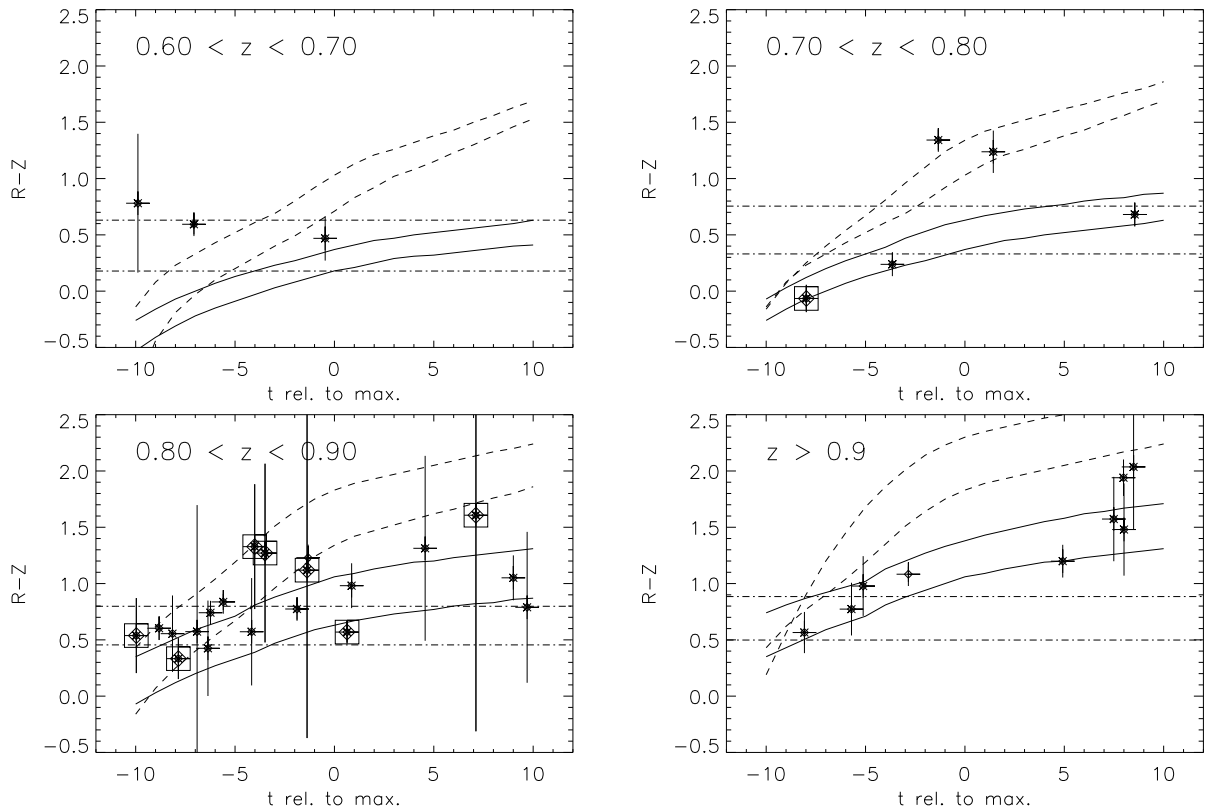}
\caption{$R-Z$ colors of the IfA Deep Survey SNe Ia as a function of
time relative to maximum brightness.  Spectroscopically confirmed SNe
Ia are indicated with a box.  Several contours are shown to indicate
the range of different types of SNe over the redshift range indicated
in each plot.  SN 1995D is a bright SN Ia (solid lines), SN 1999by is
an extremely subluminous SN Ia (dashed lines).  Type II SN
(dashed-dotted lines) are represented by extrapolation from single
measurements of SNe 1998S and 1999em, inflating the width by 0.3
magnitudes to allow for evolution.  The objects without spectroscopic
confirmation as SNe Ia are all photometrically consistent with being
so.  Except at very high redshift, they are not inconsistent with the
SN II regions, however. }
\label{likelycolors}
\end{figure}

These templates allow us to determine the expected photometric colors
as a function of time for the various classifications of supernovae.
Figure ~\ref{likelycolors} shows $R-Z$ as a function of time relative
to maximum brightness for 17 of the IfA Deep Survey SNe, divided into
redshift bins of 0.1, from $z=0.60-1.0$.  Contours indicate the
expected evolution of the two SNe Ia templates at these redshifts.
For the SNe II, we calculated $R-Z$ boundaries of each redshift range
for both of the templates, and inflated each color region by 0.15
magnitudes to allow for evolution as well as uncertainty in the true
dispersion of the population.  The times relative to maximum are
calculated using light curve fits from Section ~\ref{sec-mlcs} below.

The IfA Deep Survey supernovae, whether spectroscopically confirmed as
SNe Ia or not, are all consistent with the contours predicted by the
SN Ia templates.  Many are not inconsistent with the SNe II contours
as plotted, however, particularly those at very early times, as well
as some at low redshift.  For the very highest redshifts, the
likely SNe Ia are all far too red by a week past maximum light to be
consistent with SNe II.  Also, as expected, at higher redshifts the
SNe lie along the contours defined by the bright SN Ia 1995D, rather
than the subluminous SN 1999by.

These plots do not include contours for SNe Ib/c, due to the lack of
suitable early-time templates in the UV.  SNe Ib/c are typically much
fainter than SNe Ia (with a luminosity function as given by Richardson
et al. (2002) of M$_{B}$(Ib/c)=-18.04, compared to
M$_{B}$(Ia)=-19.46), so we do not expect to be significantly
contaminated at high redshift by these objects for the same reasons as
for the faint SN Ia 1999by.  However, SN Ic 1992ar (Clocchiatti et
al. 2000) was potentially one of the brightest supernovae of any type
ever observed, pointing out the risk in any argument based on the
luminosity function.  Richardson et al. (2002) further note that five
of the eighteen SNe Ib/c in their sample (including SN 1992ar) are as
bright or brigher than SNe Ia, possibly suggesting a bimodal
distribution of faint and bright events.

Since these fourteen SNe are all consistent with SNe Ia at the
appropriate redshift (although not necessarily inconsistent with other
types of SNe), we have decided to continue to include them in our
sample.  In Section ~\ref{sec-mlcs} we will mention a goodness-of-fit
criterion which was used to further bolster our confidence in their
inclusion.  When we perform cosmological density parameter
calculations in Section \ref{sec-anal} below, we will do so using our
entire sample of SNe discussed here, as well as with only the 9
unambiguous SNe Ia.

\subsection{Discussion of IfA Deep Survey Yield}
\label{sec-yield}

In Section ~\ref{sec-obs}, we noted that the IfA Deep Survey was
expected to discover and monitor 10--25 SNe Ia in the redshift range
of $0.9<z<1.2$.  However, here we have reported only 4 such objects,
with an additional 8 at slightly lower redshift ($0.8<z<0.9$).  This
raises a question that was the subject of much discussion even while
the survey was still in progress: Why did we find so few SNe Ia at the
highest redshifts?

The first possible answer is that we expected far too many objects due
to overestimating the rates of SNe Ia at these redshifts.  While it is
possible that previous surveys with a lower sensitivity have
overestimated the SNe rate at extremely high redshifts, there have
been enough surveys (see Tonry et al. 2003) exploring out to high
redshifts to indicate that our yield was unexpectedly and anomalously
sparse.  The continuous nature of the IfA Deep Survey, which has
allowed us to augment our yield beyond those spectroscopically
confirmed SNe Ia, means that we will be able to re-search the
observations and potentially discover supernovae that may have been
missed during the survey (for examples, SNe at the cores of galaxies
may be misclassified as AGN).  It will of course be impossible to
obtain a supernova spectrum at this time, but in many cases we should
be able to measure a host galaxy redshift, as was done with the
November 2002 observation of SN 2001jn.

There are several additional potential explanations related to the
details of the survey and its implementation.  The first of these
could be that our survey did not go as deep as we had initially
expected.  Another possible answer is that we did not do a complete
job of discovering supernovae that were present in the observations.
We did indeed discover $z\approx1$ SNe Ia (objects at this flux level
were not difficult to spot), as well as numerous other similar objects
which were not spectroscopically confirmed, so we do not feel either
of these explanations are correct.  However, in any magnitude-limited
survey a luminosity bias must be expected, so it may be that our
sample is simply not complete to redshift $z\approx1$ despite our
expectations.  There are also possible spectroscopic explanations
analogous to the above speculations.  For SNe at $z\approx1$, at the
extreme limits of what can be observed, good luck in both the timing
and conditions of spectroscopic nights are crucial for successful
observations.  These concerns with both the photometric and
spectroscopic observations illustrate the extreme difficulty in
attempting large scale surveys for $z\approx1$ SNe from the ground.
Even with regular and frequent access to telescopes with the ability
to detect such SNe, the vagaries of the atmosphere cannot be predicted
in advance.  And even when provided with advantageous weather, the
exposure times required for spectroscopic confirmation mean that only
a small fraction of discovered supernovae will be properly observed
spectroscopically.  The recent demonstration of the ACS grism on $HST$
to obtain an identifiable spectrum of a $z=1.3$ SNe Ia (Riess et
al. 2003) shows the future of $z>1$ supernova surveys is undoubtedly
in space.

\subsection{Supernova Light Curves}
\label{sec-lightcurves}

As discussed by Novicki \& Tonry (2000) and Tonry et al. (2003), and
expounded upon by Barris, Novicki, \& Tonry (in preparation), we have
developed a new method for calculating supernova light curves.  The
classic method for measuring SN fluxes is to obtain a template image,
typically either at the start of a campaign (before the supernovae to
be discovered have exploded) or at the end (often up to a year after
the conclusion of the campaign, to ensure the supernova has faded
completely), in which the supernova will not be present, so that the
result of subtracting from an image of the supernova will yield the
correct flux.  If the template actually has a low level of supernova
flux, the derived magnitudes will be incorrect.  If it is necessary to
wait up to a year to obtain a template image, there will be a
significant delay in producing results.  Furthermore, if this template
image has poor seeing or low signal-to-noise, it will create large
uncertainties in measured flux even if the supernova images themselves
are of high quality.

Our new method, dubbed N(N-1)/2, collects all observations of a given
supernova, subtracts every pair of images, and solves a corresponding
matrix of flux differences.  Novicki \& Tonry (2000) demonstrated that
this can lead to a decrease in uncertainties by a factor of
$\sqrt{2}$, due to effectively using every image as a template, thus
eliminating dependence upon a single exposure.  Since there is no
pre-defined zero-flux template in the N(N-1)/2 method, there is an
ambiguity in the flux zero-point, which creates interesting issues for
photometric fitting of SNe that are discussed more in Section
~\ref{sec-mlcs}.

The light curves for our 23 SNe as calculated via the N(N-1)/2 method
are given in Table ~\ref{table:baryonyx-observations}.  These tables
include the date of each observation and the measured flux of the
supernovae, as well as information from the fits of the data,
described below.  All flux values are scaled so that flux=1
corresponds to $m=25.0$, so that magnitudes may be calculated by
$$m=-2.5 \log(flux) + 25.0.$$
This is of course not properly defined for values of flux $<0$, which
indicate the lack of a detection in the given observation.

Uncertainties in magnitudes may be calculated directly from the
uncertainties in flux according to the formula
$$\sigma(mag) = 2.5 \log ( 1 + \sigma(flux)/flux ),$$
although these values will not be precise, as the uncertainties in
magnitude are not symmetric due to the logarithm operation.

\subsection{$HST$ Photometric Observations}
\label{sec-hst-obs}

Our $HST$ Cycle 10 allocation allowed us to obtain light curves of
several supernovae with WFPC2 using the $F850LP$ filter.  Since the
IfA Deep Survey involved 5 different field positions, we could not
predict in advance the location of the best supernova candidate(s) for
each month.  The Telescope Time Review Board approved our request to
change our observations to Targets of Opportunity (ToO), which allowed
us to pick the best supernova candidate, regardless of position.

After spectroscopically observing our candidates and confirming their
identity as SNe Ia at a desirable redshift, selected objects were sent
to $HST$ for observation.  The process of discovery, spectroscopic
analysis, and notification to STScI of the ToO targets creates an
unavoidable time gap of about 10--12 days between the discovery and
the first $HST$ observations.  Typically, the discovery epoch of a
high-$z$ supernova is a few days before maximum brightness, and
although the time dilation factor of ($1+z$) works to lessen the delay
in the rest frame, none of our $HST$ light curves begins until past
maximum light.  We observed SNe 2001hu, 2001jf, and SN 2001jh, with
relevant information given in Table ~\ref{table:hstobs}.

Each SN observational epoch consisted of approximately 3--5
orbits. The data were combined using the drizzle procedure outlined by
Koekemoer et al. (2002).  Determining accurate photometry from the
WFPC2 images requires properly correcting for various CCD and optical
effects, most importantly the non-unity charge transfer efficiency
(CTE).  We followed the procedure outlined by Dolphin (2000) in order
to measure reliable PSF-fitting photometry.  Along with the magnitude
of the supernova, a few nearby stars were measured with the same
photometric method.  There were 5 epochs for SN 2001jh and 6 each for
SNe 2001hu and 2001jf, so the same stars were measured several times
with consistent results.

We also obtained ACS images well after the completion of the survey to
serve as templates in order to subtract in the same manner as for the
ground-based observations.  These subtractions were not done using the
N(N-1)/2 method, but rather used the classic single-template method.
This was done for the sake of simplicity, but should be acceptable
since none of the problems associated with the single-template method
is an issue (poor seeing, S/N, or timing of the template).

\section{Distance Determination}
\label{sec-dist}

\subsection{The Multi-wavelength Light Curve Shape Method}
\label{sec-mlcs}

In order to use these SNe Ia for cosmological analysis, we created a
new version of the Multi-wavelength Light Curve Shape (MLCS) analysis
method (see Riess, Press, \& Kirshner (1996, hereafter RPK96), and Jha
2002).  This new implementation was developed in consultation with
authors of previous versions of the MLCS fitting code, and features
few substantial changes from them.  The MLCS method simultaneously
fits for distance modulus ($m-M$), $A_{V}$, and $\Delta$, a parameter
defined by the difference in absolute magnitude between a given
supernova and a fiducial SN Ia.  This $\Delta$ parameter therefore
describes the shape of the SN light curve, since there is a
correlation between absolute magnitude and light curve shape (see
Phillips 1993).

We first constructed MLCS templates through iteratively fitting a
sample of 32 low-$z$ SNe Ia taken from the Cal\'{a}n/Tololo survey
(Hamuy et al. 1996), as well as from Riess et al.  (1999b) and Jha
(2002).  At the end of this training process the fits to these 32 SNe
Ia produced a scatter of 0.14 magnitudes around the Hubble Diagram.

K-corrections were calculated using the formulae described in Kim,
Goobar, \& Perlmutter (1996) and Schmidt et al. (1998), using a set of
135 SN Ia spectra ranging from 14 days before maximum light to 92 days
after maximum light. As described by Nugent, Kim, \& Perlmutter (2002)
as well as Germany et al. (2003), before applying the K-correction
formulae, the SN Ia spectra are first matched to the $B-V$ color of
the MLCS template by applying the Savage-Mathis (1979) reddening law.
The Schlegel, Finkbeiner, \& Davis (1998, hereafter SFD) galactic
extinction is applied to the Ia spectra set, and the spectra are then
stretched by the appropriate factor of ($1+z$).  These modified
spectra are used to calculate the K-correction, providing a series of
K-correction estimates as a function of SN age, which is fit with a
3-knot spline. The resulting K-corrections are interpolated from this
spline, and the uncertainty in the K-corrections estimated from the
scatter about the fitted relationship.  These K-corrections are then
used to fit the best template (which has its own $B-V$ evolution based
on its intrinsic colour and fitted extinction), with this new $B-V$
evolution used in place of the MLCS template, and the subsequent steps
repeated until convergence is reached. Typically, this iterative
process changes the K-corrections by $<$0.03 mag.

Calculating proper K-corrections depends upon knowing the shape of the
light curve (i.e. the MLCS $\Delta$ parameter) as well as the
reddening of the supernova spectrum ($A_{V}$).  Thus in the fitting
procedure we were forced to select given input values for $\Delta$ and
$A_{V}$, determine the necessary K-corrections for such a light curve
shape, and then fit for the best output set of parameters.  If the
best-fit values of $\Delta$ and $A_{V}$ were equal to the input values
(a rigid constraint of within 0.01 for each parameter was used), then
the solution was deemed acceptable.  During the fit procedure, the
constraint that $A_{V} > 0$ was applied, rather than allowing for a
solution with negative extinction.  This eliminates the need to apply
a Bayesian prior after the fact (as done by RPK96), which affects
$A_{V}$ while leaving untouched the other, correlated, parameters
($m-M$) and $\Delta$.

For each MLCS run on a supernova, the appropriate SFD galactic
extinction is applied to the light curve (for the $Z$ filter used
here, we calculated a value of $A/E_{B-V}$ of 1.520 following the
description from SFD), the points are shifted so that $t=0$ at
$t=t_{0}$, and the time dilation factor of ($1+z$) applied.  For
high-redshift objects ($z>0.7$), we K-corrected $I$ band to $B$, and
$Z$ to $V$, to match with the filters we had used to train the MLCS
method.  For SNe 2001hu, 2001jf, and 2001jh we also K-corrected
$F850LP$ to $V$.  For low-redshift objects ($z<0.7$) we shifted $R$
and $I$ bands to $B$ and $V$, respectively.  Tables
~\ref{table:baryonyx-observations} and ~\ref{table:hstobs} include
K-correction values for the filters used in MLCS fitting.  Our MLCS
templates extended from 10 days prior to maximum light in the rest
frame to 40 days post-maximum.  The reason for this late-time cutoff
was that the templates were created with this survey in mind, and so
were tailored for SNe Ia at $z\approx1.0$, which will fade from
visibility by this point in the light curve, rather than for lower
redshift objects, which will still be visible.  The early-time cutoff
is due to the paucity of low-$z$ supernova observations at earlier
epochs for constructing MLCS templates.  However, we have extended our
fits to earlier times using a prescription similar to that of Riess et
al. (1999a), with a slight modification to account for zero-point
differences.  This extension allows us to take advantage of the
unprecedented number of very early light curve points provided by the
continuous nature of the survey.

The use of the N(N-1)/2 method for producing light curves did lead to
some complications in the light curve analysis due to the ambiguity
concerning the flux zero point.  The proper way to account for this is
to recognize that the flux zero point is in fact a free parameter, and
fit for it accordingly.  Thus, our MLCS code, in addition to fitting
for the time of maximum $t_{0}$ (externally, through iteration),
distance modulus $(m-M)$, extinction $A_{V}$, and MLCS delta parameter
$\Delta$ (constrained to lie between $-0.6 < \Delta < +0.6$), also
fits for offsets in each filter, $\delta_{B}$ and $\delta_{V}$.  These
offsets are done in flux, not magnitude, since these are flux
differences rather than a multiplicative factor.

A further constraint was placed on satisfactory values of $\delta_{B}$
and $\delta_{V}$.  We can ascertain that our zero points are roughly
reliable since our time baseline is long enough that, in practice,
there is always an observation with little or no supernova flux.
Because of our confidence in the general accuracy of our flux zero
point, we only accept solutions with small values for $\delta_{B}$ and
$\delta_{V}$.  Also, for SNe 2001hu, 2001jf, and 2001jh, we did not
fit for a flux offset for the $F850LP$ points, as these magnitudes
were calculated using the single-template method.

\begin{figure}
\epsscale{2.2}
\plotone{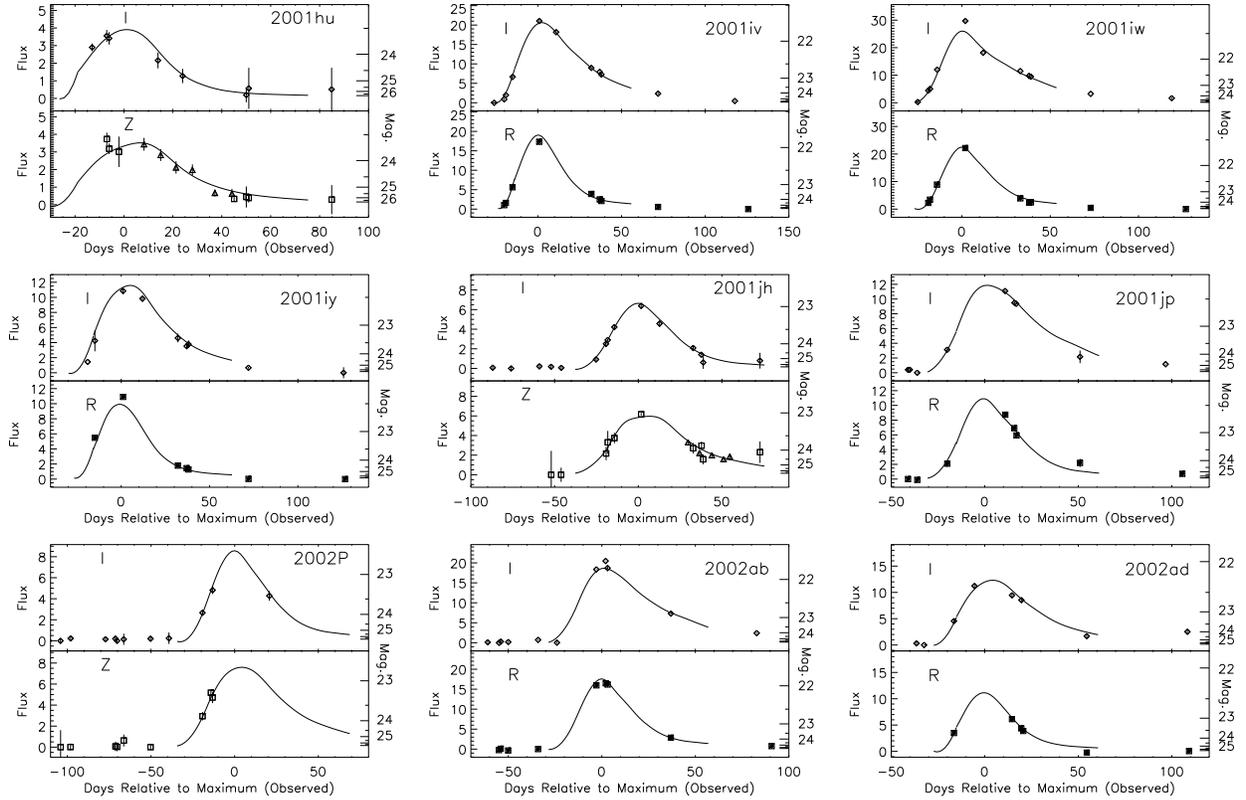}
\caption{Light curves for the 9 IfA Deep Survey SNID-confirmed SNe Ia.
Ground based data are plotted as observed for the two filters fit with
MLCS ($R$,$I$ for $z<0.7$ and $I$,$Z$ for $z>0.7$), with the MLCS fit
shown by solid lines.  In addition, $F850LP$ (triangles) points for SN
2001hu and 2001jh are shown with a shift equal to the difference in
K-corrections from $Z \rightarrow V$ and $F850LP \rightarrow V$ to
better illustrate the fit to the data.}
\label{snialc}
\end{figure}

\begin{figure}
\epsscale{2.2}
\plotone{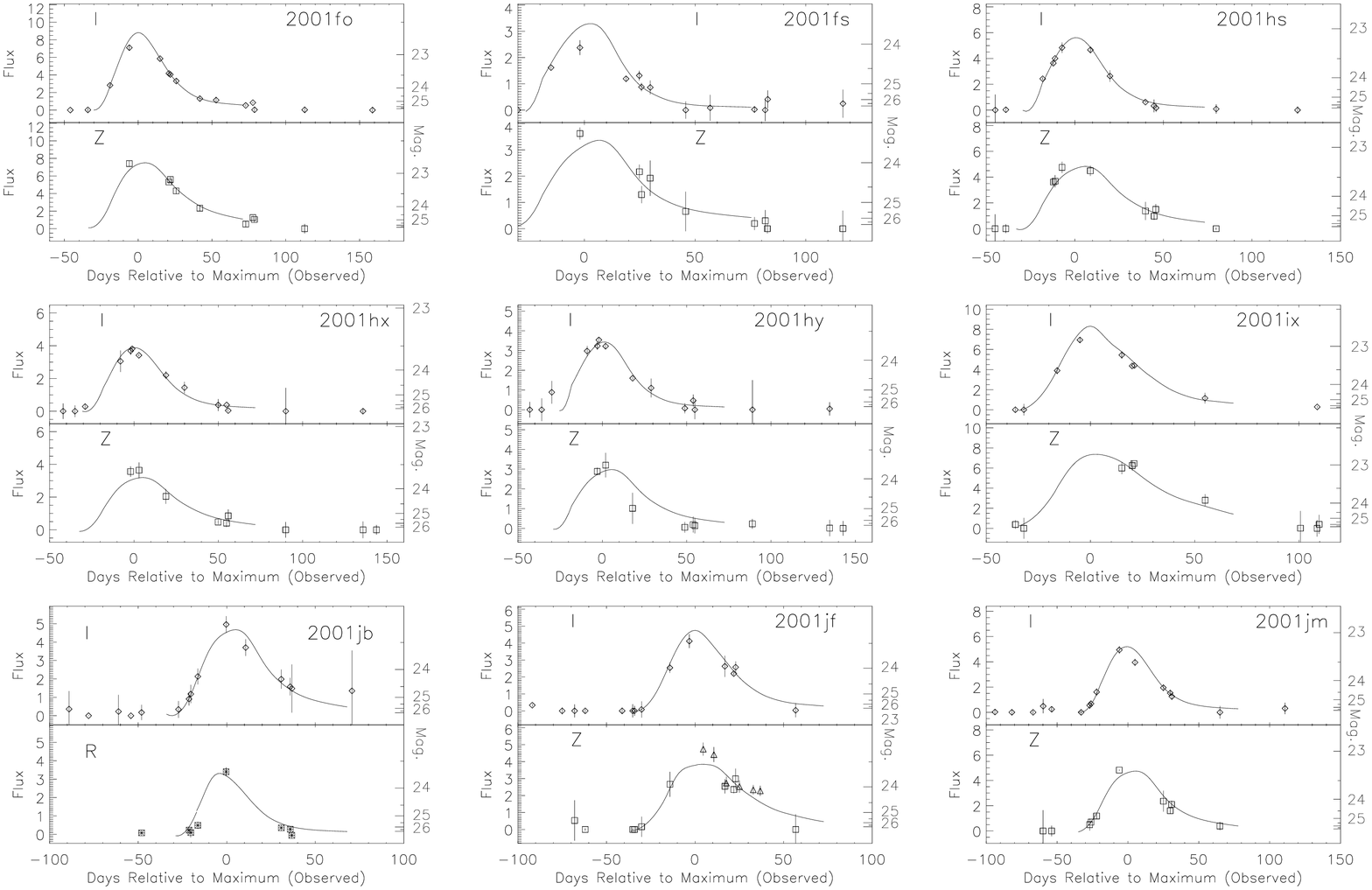}
\label{lc-more}
\end{figure}

Best-fit MLCS parameters ($m-M$, $A_{V}$, $\Delta$) for the IfA Deep
Survey SNe Ia are given in Table ~\ref{table:mlcsinfo}.  Table
~\ref{table:baryonyx-observations} includes the values for time
relative to $B$ band maximum in the supernova rest frame for every
observation.  Uncertainties in the parameters ($m-M$, $A_{V}$,
$\Delta$) were calculated in the same way as given in RPK96.  Light
curves with MLCS fits are shown in Figure ~\ref{snialc} for the 9
spectroscopically confirmed SNe Ia and Figure ~\ref{lc-more} for the
14 additional SNe Ia.  In these figures, we have plotted the flux
values for all SNe scaled so that $magnitude=25$ corresponds to a
value of $flux=1$.

We have also included in Table ~\ref{table:mlcsinfo} the $\chi^{2}/N$
values for the MLCS fits for all 23 SNe.  The fits are quite good for
all of the SNe Ia, both spectroscopically confirmed with SNID and
otherwise.  These $\chi^{2}/N$ values for the 14 likely SNe Ia lend
further confidence to their identification as SNe Ia and inclusion in
our sample.

\subsection{Additional Fitting Methods }
\label{sec-otherfits}

In addition to MLCS fits, we also analyzed the SN Ia light curves with
the Bayesian Adapted Template Match (BATM) Method (Tonry, in
preparation) and dm15 (Germany 2001).  BATM uses a set of
approximately 20 well observed nearby supernova light curves in
combination with $\sim$100 observed spectral energy distributions
(SEDs) in an attempt to span the expected behavior of SNe Ia.  For all
pairs of light curve and SED, BATM calculates likelihoods as a
function of distance $d$, extinction $A_{V}$, and time of explosion
$t_{0}$, further marginalizing over $t_{0}$ and applying a prior on
$A_{V}$ as described by Tonry et al. (2003), in order to measure $d$
and $A_{V}$.  The dm15 method is a modification of the
$\Delta$m$_{15}$ method of Phillips et al. (1999), which utilizes the
facts that the decline in magnitudes of SNe Ia in the first 15 days
after $B$-band maximum light correlates with luminosity, and that the
late-time color curves of all unreddened SNe Ia are uniform,
regardless of decline rate (Lira 1995; however, the peculiar SN Ia
2002cx (Li et al. 2001), does not follow the standard Lira relation).
In the dm15 method, a set of 15 template light curves of nearby SNe Ia
is used to measure distance $d$, marginalizing over $t_{0}$ and
$A_{V}$ for each template with an acceptable fit.

Results of light curve fits for the IfA Deep Survey SNe Ia from each
of these two methods are given in Table ~\ref{table:otherfitinfo} (The
dm15 method was not used for SNe at $z>0.8$ due to the lack of
observed filters to match its templates).  The three methods are all
quite consistent with each other, with scatter between each pair of
methods of $\sim$0.1 magnitudes.  We have also combined all our
measurements into a single value for distance in the same manner as
Tonry et al. (2003).  Zero-point differences between each method were
computed by comparing common SN measurements, distances placed on a
Hubble flow zero-point ($dH_{0}$), and the median selected as the best
distance estimate.  Uncertainty in the distance estimate was taken
from the median of the error estimates from the individual methods,
scaled down by the $1/4$ power of the number of contributors.  As
described by Tonry et al., this median procedure is not expected to
significantly improve the accuracy of the final estimate, but ideally
will result in a more Gaussian-like distribution.  Supernova
properties derived by combining results from all the methods are
included in Table ~\ref{table:summarytable} in a form designed to be
similar to those given by the summation of Tonry et al. (2003).

\section{Cosmology with IfA Deep Survey Supernova}
\label{sec-anal}

Tonry et al. (2003) collected redshifts and distances for all
published SNe Ia at cosmological distances.  Whenever possible, they
also performed various light curve fits to the published photometry in
order to place as many as possible on a common system.  They present
redshift and luminosity distance for a total of 230 SNe Ia.  However,
this includes many objects which may be unsuitable for cosmological
analysis, particularly those which are heavily extinguished or are
nearby enough for velocity uncertainties to be a major problem.

The results from the distance fits from the previous section are
illustrated on a Hubble diagram in Figure ~\ref{hubble195}, which
includes all of our SNe Ia, as well as those from Tonry et al. (2003).
These figures show the result of subtracting the distance modulus
predicted from an ``empty universe,'' i.e. a cosmology with
($\Omega_{M},\Omega_{\Lambda}$)=(0.0, 0.0), from that measured for
each supernova.  Although the distances given in Tables
~\ref{table:mlcsinfo} and ~\ref{table:otherfitinfo} are in the form of
($m-M$) for the sake of familiarity, to construct the figures we only
need use the values from Table ~\ref{table:summarytable}, which have
no dependence on $H_{0}$.  Immediately obvious on the plot is the
large number of literature points at $z\approx0.5$ with a positive
deviation, which is the signature of an accelerating universe and a
cosmological constant-like term.  Supernova surveys targeting
$z\approx0.5$ were well placed to detect the presence of
$\Omega_{\Lambda}$.  A survey for supernovae at $z=1.0$, on the other
hand, is designed to target the redshift region where the deviation
between a $\Omega_{\Lambda}$-dominated universe and a systematic
effect proportional to redshift is large.  This was the goal of the
IfA Deep Survey, though it turned out that we discovered supernovae
over a fairly large range in redshift, and many fewer than expected at
$z\approx1$, as was discussed in Section ~\ref{sec-yield}.
Nevertheless, the number of $z\geq0.7$ SNe (15, see Table
~\ref{table:summarytable}) which the IfA Deep Survey has added to what
was heretofore an extremely sparse region of redshift space is still
substantial, doubling previously published results (12 from the
collection presented by Tonry et al. 2003, plus 3 from Knop et
al. 2003 which are not included in this analysis).

We have also taken medians by redshift bins in order to better
illustrate the overall trend with redshift, using the subset of 200
literature SNe with $A_{V}\leq0.5$ plus 22 from this survey, and
requiring that bins must have a width of at least 0.25 in log $z$, and
contain at least 20 SNe.  Results are shown in Figure
~\ref{hubblemedian}.  Our new SNe have continued to fill in the
highest redshift bins, and show an even more rapid and sharp turnover
than was observed by Tonry et al.  The median magnitude deviation
relative to an empty universe for the highest-$z$ bin, calculated as
0.00 magnitudes in Tonry et al. for 12 objects centered at $z=0.87$,
is now $-0.07$ magnitudes for 20 objects centered at $z=0.89$, while
the uncertainty in this bin, estimated by the 68\% scatter of points
in the bin, is 0.07 mags compared to 0.08 mags from the Tonry et
al. sample.  These changes are due to the fact that the SNe at
$z>0.85$, which are predominantly from this survey, are overwhelmingly
$\it brighter$ than the empty-universe cosmology, compared to those at
slightly lower redshift which tend to be fainter than this model.
Although the region at $z\approx1$ is still underpopulated, it is
becoming more and more apparent from the sample presented here that
the trend of the population of SNe Ia at $z>0.5$ is more consistent
with the turnover predicted by an $\Omega_{\Lambda}$-dominated
cosmology rather than a systematic effect which increases with
redshift (see Figure ~\ref{hubblemedian}).

\subsection{Cosmological Density Parameter Determination}

We now turn to determination of cosmological density parameters from
the supernova data.  We wish to calculate $\chi^{2}$ as a function of
the parameters ($H_{0},\Omega_{M},\Omega_{\Lambda}$).  Since we have
SN Ia distances over a wide range of redshift, we are able to
marginalize over $H_{0}$ and concentrate solely on the $\Omega$
parameters.  $H_{0}$ appears as a quadratic term in $\chi^{2}$, as
shown below, so it appears as a separable Gaussian factor in the
probability to be marginalized over, and doing so over $H_{0}$ is
equivalent to evaluating $\chi^{2}$ at its minimum with respect to
$H_{0}$.

The distance modulus for the observations may be rewritten as
$$(m-M)_{obs} = 5 \langle\log(dH_{0})\rangle + 25 - 5\ \log H_{0,1}.$$
For a given cosmology, the luminosity distance can be expressed as 
$$d_{lum}=f(z,\Omega_{M},\Omega_{\Lambda})/H_{0,2} .$$
Therefore
$$(m-M)_{model} = 5\ \log f + 25 - 5\ \log H_{0,2} ,$$
where the Hubble parameter $H_{0,2}$ used for calculating the model is
not necessarily equal to that used for the observational data
($H_{0,1}$).  This would not be by choice, but rather reflects the
fact that the distances presented by Tonry et al. (2003), which are
tied to the Hubble flow, are related to the ``true'' $H_{0}$ only by a
fit parameter and therefore any new fit of distance must revisit this
parameter, necessitating marginalization over $H_{0}$.

Calculating $\chi^{2}$ is simply the sum over all supernovae

$\chi^{2} = \sum \frac{\displaystyle((m-M)_{obs} -
(m-M)_{model})^{2}}{\displaystyle\sigma(\langle\log(dH_{0})\rangle)^{2}} $

$$\chi^{2} = \sum \frac{\displaystyle(5 \langle\log(dH_{0})\rangle - 5\ \log f + 5\
\log(H_{0,2}/H_{0,1}))^{2}}{\displaystyle\sigma(\langle\log(dH_{0})\rangle)^{2}} $$
where, as in Tonry et al. (2003), we also add a 500 km sec$^{-1}$
uncertainty in quadrature to the redshift errors given therein to
account for uncertainties due to peculiar motions.

For any cosmology ($\Omega_{M},\Omega_{\Lambda};z$), $f$ is a given,
and the ratio $H_{0,2}/H_{0,1}$ is an unknown parameter on which
$\chi^{2}$ depends.  The minimum value of $\chi^{2}$ occurs where this
parameter has the value

$log(H_{0,2}/H_{0,1})_{min} = - \frac{\displaystyle\sum
\frac{\displaystyle5 \langle\log(dH_{0})\rangle - 5\ \log
f}{\displaystyle\sigma(\langle\log(dH_{0})\rangle)^{2}}}{\displaystyle\sum
\frac{\displaystyle 1}{\displaystyle\sigma(\langle\log(dH_{0})\rangle)^{2}}} .$

We then convert to a probability value proportional to $\exp(-0.5
\chi^{2}$), with $\chi^{2}$ evaluated at this minimum, which allows us
to determine contours of constant probability density for
($\Omega_{m}$, $\Omega_{\Lambda}$).

We can now add our new high-redshift supernovae to the previously
published SNe and investigate the implications for cosmology.  For
determination of cosmological parameters, Tonry et al. (2003) used
cutoffs of $z > 0.01$ and $A_{V} \leq 0.5$, and we have chosen to
adopt the same restrictions, leaving a sample of 172 objects.  A first
test is to calculate cosmological parameter best-fit regions using
only the IfA Deep Survey SNe.  The low-$z$ dataset is crucial to
constrain the Hubble constant, so we include objects with $z\leq0.30$,
but no objects at higher redshift save the 22 IfA Deep Survey SNe with
$A_{V} \leq 0.5$ (eliminating SN 2001jn, as seen by Table
~\ref{table:summarytable}).  This reduces the sample size to 98 from
the compilation by Tonry et al. (2003) and 22 from this study, for a
total of 120 objects.  With these 120 SNe Ia, we obtain a best-fit
value of $\chi^{2}$=97.7 at ($\Omega_{m}$, $\Omega_{\Lambda}$)=(0.27,
0.36).  Corresponding probability contours derived from this sample
are shown in Figure ~\ref{contours}a.  The best-fit value with
$\Omega_{total} = 1.0$ (consistent with measurements of the cosmic
microwave background (CMB), see below), is ($\Omega_{m}$,
$\Omega_{\Lambda}$)=(0.42, 0.58).  If we further restrict our high-$z$
sample to only those 9 SNe Ia for which we obtained unambiguous
spectral SN Ia confirmation, for a total set of 107, the minimum value
of $\chi^{2}$=85.2 occurs at ($\Omega_{m}$, $\Omega_{\Lambda}$)=(0.09,
0.15) (see Figure ~\ref{contours}b).  The fact that $\chi^{2}/N$ is
much less than 1 is due to the 98 SNe Ia at $z\leq0.30$, whose
contribution to $\chi^{2}$ for an empty-universe is 77.3.  The 9 IfA
Deep Survey SNe identified as SNe Ia by SNID contribute approximately
8 units of $\chi^{2}$, with the additional 13 SNe Ia contributing
approximately 12 units, indicating both that there is no sign of a
difference in the distributions of the two subsamples as might be
feared if there were incorrectly identified SNe Ia in the latter, and
that our distance uncertainties are reasonable.

The fact that the IfA Deep Survey SNe agree well with an empty-universe
cosmology is to be expected based on their redshift distribution.  As
is seen by the models plotted in Figure ~\ref{hubble195}, in the
redshift range $z=0.8--1.0$ the difference between an
$\Omega_{\Lambda}$-dominated universe and an empty-universe is
decreasing, meaning there is little power to differentiate between
these models from such SNe.  

Adding our 22 SNe Ia to the full Tonry et al. dataset of 172 objects
and following the same procedure, we obtain the contours shown in
Figure ~\ref{contours}c.  The best-fit value is $\chi^{2}$=195.5 at
($\Omega_{m}$, $\Omega_{\Lambda}$)=(0.71, 1.28).  These contours are
quite similar to those obtained by Tonry et al., who found a best fit
at ($\Omega_{m}$, $\Omega_{\Lambda}$)=(0.69, 1.34), which is to be
expected since the new data are statistically compatible with the
Tonry et al. set.  If we add the constraint that $\Omega_{total} =
1.0$, we obtain a best-fit at ($\Omega_{m}$,
$\Omega_{\Lambda}$)=(0.33, 0.67).  It is interesting to note, as did
Tonry et al., that the supernova results intersect the line
$\Omega_{total} = 1.0$ at the point which is also consistent with
constraints from the 2dF survey of $\Omega_{M} h = 0.20 \pm 0.03$.

The best-fit value for the sample of 194 SNe Ia, however, is far from
the $\Omega_{total} = 1.0$ line, although it is consistent with such
values.  The IfA Deep Survey SNe contours, while larger due to smaller
sample size, are centered closer to this line.  The majority of
objects at $z\approx0.5$ lie $above$ the ($\Omega_{m}$,
$\Omega_{\Lambda}$)=(0.3, 0.7) model, pushing $\Omega_{\Lambda}$
higher, while those at $z>0.7$ mostly lie $below$ the line, which
pushes $\Omega_{M}$ higher in order to compensate, moving the contours
into the region seen in Figure ~\ref{contours}c.  The reason can be
seen from the median values given in Figure ~\ref{hubblemedian}, where
the population of SNe at z$\approx$0.4-0.6 becomes fainter than those
at lower redshift, only to have an extreme drop again for the highest
redshift bin.  These subsamples are drawn from many different sources,
and within any previous study there was often a very heterogeneous mix
of observations.  For example, the light curves presented by Tonry et
al. (2003) were obtained on as many as 6 different
telescope/instrument combinations, even within a single filter,
introducing many possibilities for photometric errors.  By contrast,
the uniformity of the observations and reductions of the IfA Deep
Survey should minimize such possible sources of bias, while still
subject to systematic errors such as uncertainties in K-corrections.
Another possible source of systematic errors involve our dependence on
the $Z$ filter, whose photometry is ill-defined compared to other
filters.

We can also explore the possible effects of gravitational lensing on
our results.  In order to do so, we perform the above calculations
with Dyer-Roeder (DR) distance formulae (Fukugita et al. 1992; see
also Dyer \& Roeder 1972, 1973), which allow for the fact that the
space through which light is propagating is inhomogeneous.  In the
above calculations, we have used the standard Robertson-Walker metric,
which treats the distribution of material in the universe as being
completely smooth.  Depending on the true distribution, this may
introduce errors in our cosmological parameter determination, as
demonstrated by Kantowski (1998).  An alternative model to consider is
one in which light from the supernovae travels along a path entirely
devoid of matter, hence the term ``empty beam,'' in contrast to
``filled beam'' for the standard model.  There are also intermediate
possibilities to consider, though we will limit ourselves to the
empty-beam alternative.

As shown by Holz (1998), an empty-beam model tends towards higher
values of $\Omega_{m}$ and lower values of $\Omega_{\Lambda}$.  The
likelihood contours also become greatly elongated, particularly
towards higher values of $\Omega_{m}$.  This effect is seen in the
comparison of our contours from Figure ~\ref{contours}c and those from
an empty-beam calculation, shown in figure ~\ref{contours}d.  Although
these contours extend to extremely large values of ($\Omega_{m}$,
$\Omega_{\Lambda}$), it is clear that adopting the empty-beam model
still requires a non-zero value of $\Omega_{\Lambda}$.

However, there is also the potential for magnification due to
gravitational lensing to affect our results.  The probability for
significant magnification of sources at $z\approx1$ are low (see
Barber et al. 2000), but when combined with potential systematic
effects, such as a selection bias for such brightened objects, could
become important and partially account for the distribution of our
$z>0.8$ SNe Ia, which lie brighter than the favored ($\Omega_{m}$,
$\Omega_{\Lambda}$)=(0.3, 0.7) model.

\subsection{Comparison with WMAP}
\label{sec-wmap}

Recent results from high angular resolution observations of the CMB
with WMAP (Spergel et al. 2003) have produced extremely tight regions
of acceptable cosmological parameters which we can compare with our
derived values.  They find that the universe is consistent with being
flat ($\Omega_{total}=1.02\pm0.02$).  All plots in Figure
~\ref{contours} indicate the line $\Omega_{total}=1.0$.  As was noted
in the previous section, the 1-$\sigma$ uncertainties for the full set
of 194 SNe Ia no longer overlap with this line.  The contours from the
IfA Deep Survey are larger, since they are based on a smaller sample,
but are centered at values in better agreement with those preferred by
the CMB.  As mentioned above, it is expected that the best-fit
contours for our SNe should lie closer than the full literature sample
to an empty-universe model, based on redshift distribution, though the
improved agreement with the CMB may also be due to the homogeneous
nature of our sample minimizing systematic effects, which would bode
well for future large-scale surveys which will have similar (though
much larger) datasets.

\section{Conclusion}
\label{sec-conclusion}

We have described in detail the observational strategy and data
reduction of the IfA Deep Survey as undertaken by a collaboration of
IfA astronomers in late 2001 and early 2002, as well as the supernova
search component carried out by the High-$z$ Supernova Search Team.
This survey has already served as an unprecedented photometric dataset
for continuous detection and follow-up of high redshift supernovae, of
which over 100 candidates were discovered (Barris et al. 2001, 2002),
including the 23 SNe presented here.  Preliminary analysis of survey
data has also already yielded numerous substellar objects (Liu et
al. 2002; Graham 2002; Mart\'\i n et al., in preparation), indicating
that large numbers of such objects will be discovered with more
detailed inspection of the data.  Similarly, the photometric dataset
produced by this survey has great promise for many areas of research
such as AGN studies, galactic structure, and galaxy clustering.  The
survey also anticipates even more ambitious future projects which will
repeatedly image large patches of sky over extended periods of time,
such as PanSTARRS, LSST, and the proposed SNAP.

\begin{figure}
\epsscale{2.2}
\plotone{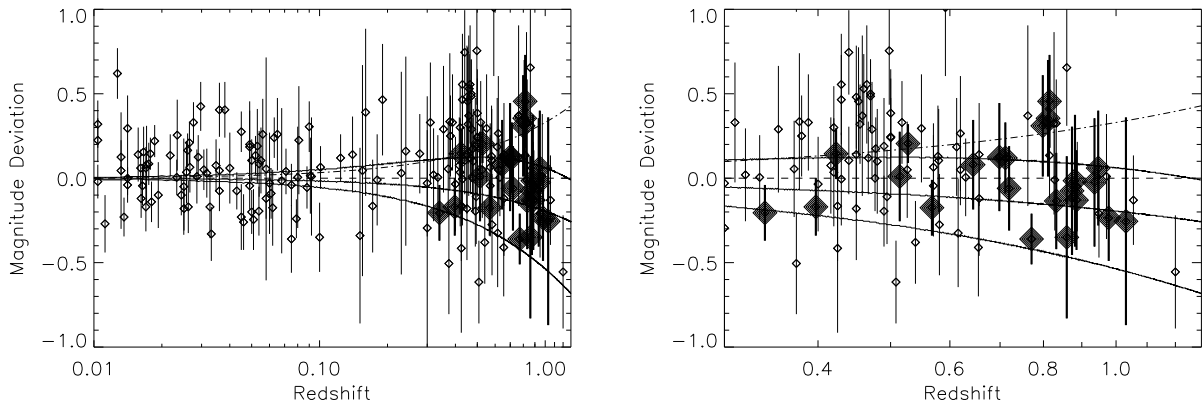}
\caption{Hubble Diagram showing the magnitude deviation with respect
to an empty universe of the 23 IfA Deep Survey SNe Ia (large
diamonds), as well as previously published SN Ia at similar redshift
collected by Tonry et al. (2003).  This diagram is constructed from
the values given in Table ~\ref{table:summarytable}, and requires no
assumption about the value of $H_{0}$.  From top to bottom, solid
lines represent cosmologies with
($\Omega_{M}$,$\Omega_{\Lambda}$)=(0.3, 0.7), (0.3, 0.0), and (1.0,
0.0), respectively.  Note the maximum positive deviation of a
$\Omega_{\Lambda}$-dominated universe occurs at $z \approx 0.5$.
Right-hand panel is focussed on high
redshift to more clearly show the IfA Deep Survey SNe.  Also shown is
a dashed-dotted line showing the effects of a systematic effect
proportional to $z$.}
\label{hubble195}
\end{figure}

\begin{figure}
\epsscale{1.0}
\plotone{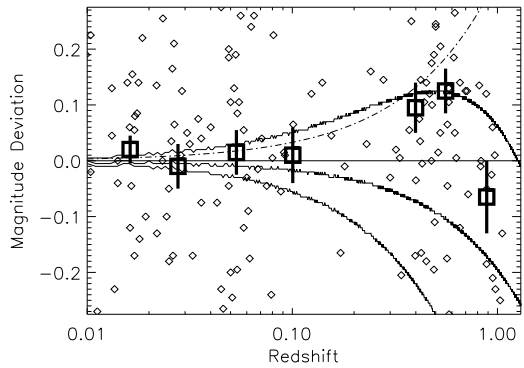}
\caption{Literature supernovae (diamonds) shown along with median
values binned by redshift (squares).  Individual points are shown
without error bars for the sake of clarity.  Redshift bins contain a
minimum of 20 SNe Ia.  From top to bottom, solid lines represent
cosmologies with ($\Omega_{M}$,$\Omega_{\Lambda}$)=(0.3, 0.7), (0.3,
0.0), and (1.0, 0.0), respectively.  Also shown is a dashed-dotted
line representing a systematic effect proportional to $z$, which does
not correspond well with the highest redshift bins. }
\label{hubblemedian}
\end{figure}

\begin{figure}
\epsscale{2.2}
\plotone{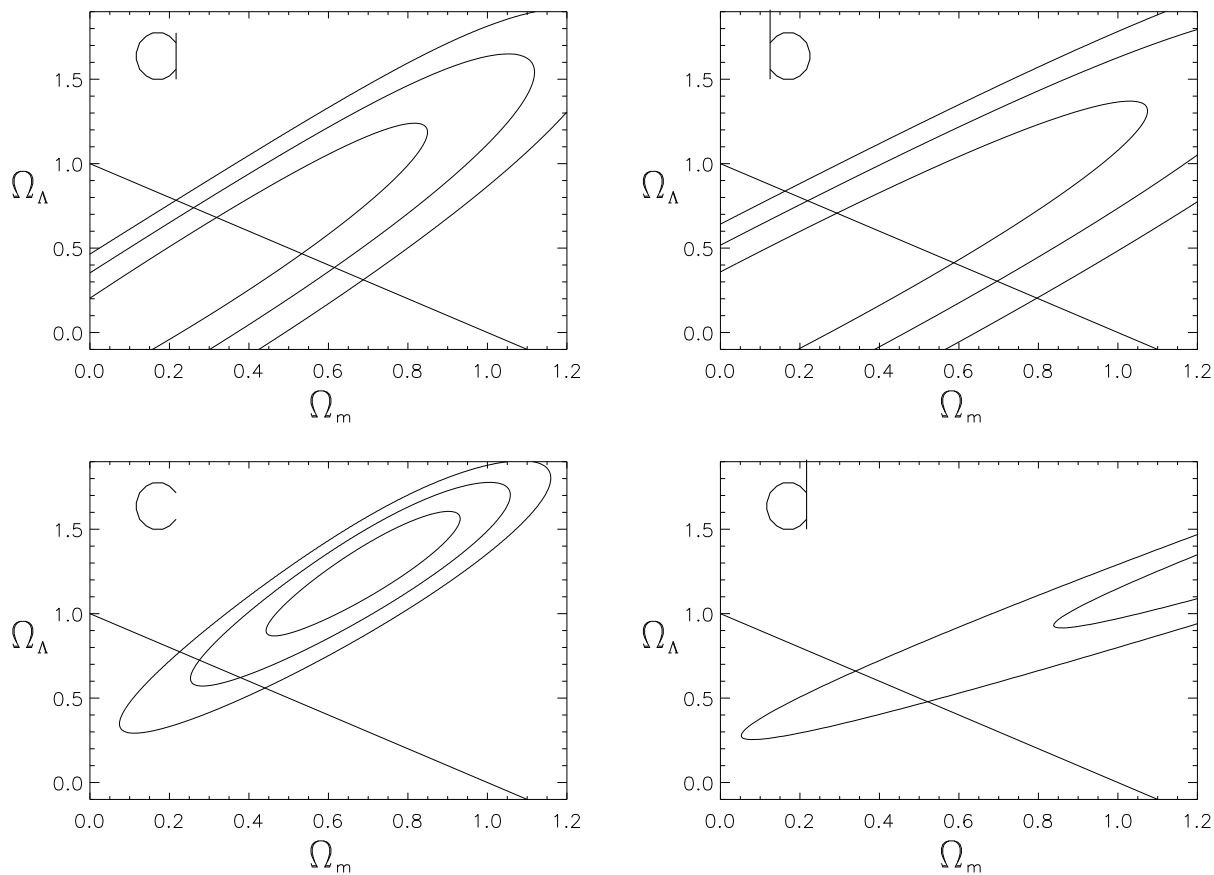}
\caption{68\%, 95\%, and 99.5\% confidence contours for the sample of:
a) 98 SNe Ia $0.01 \leq z\leq$0.3 SNe Ia plus 22 from this survey (120
total);
b) 98 SNe Ia $0.01 \leq z\leq$0.3 SNe Ia plus 9 from this survey with
SNID confirmation (107 total);
c) 172 SNe Ia with $z\geq$0.01 SNe and $A_{V}\leq0.50$ from Tonry et
al. 2003 plus 22 from the IfA Deep survey (194 total);
d) same sample as c., calculated with
the Dyer-Roeder empty-beam cosmology.  These contours are shifted to
lower values of $\Omega_{\Lambda}$ and higher values of $\Omega_{M}$,
with the trend becoming more pronounced at higher values of
$\Omega_{M}$, thus greatly elongating the contours.}
\label{contours}
\end{figure}

These 23 SNe include 15 which double the previously published sample
size of $z>0.7$ supernovae.  This region of redshift space is
extremely important for distinguishing between systematic effects and
cosmological evolution.  These supernovae, in combination with the
published body of SNe Ia, do not show evidence for continuing to grow
ever fainter at higher values of $z$, as would be expected by a
systematic effect proportional to redshift (see Figures
~\ref{hubble195} and ~\ref{hubblemedian}).  We have performed
cosmological density parameter fits using different subsets of the 23
SNe---the sample of 9 objects which are unambiguously
spectroscopically identified as SNe Ia, and the sample of 22 which
have $A_{V}\leq0.5$ (see Table ~\ref{table:summarytable}).  Both
samples are consistent with the geometrically flat universe preferred
by studies of the CMB (Figure ~\ref{contours}).  With the constraint
of $\Omega_{total} = 1.0$, we obtain a best-fit at ($\Omega_{m}$,
$\Omega_{\Lambda}$)=(0.33, 0.67) using our set of 22 and the
literature collection presented by Tonry et al. (2003).  Future
studies which will produce similarly homogeneous datasets on an even
larger scale may continue to show better agreement with the CMB and
other constraints on cosmological density parameters, as our subsample
does compared to the full literature sample.

Our yield, though impressive, was smaller than anticipated,
demonstrating the difficulty of successfully finding SNe Ia at
z$\approx$1 and higher from the ground, even with a well-planned and
executed survey using some of the world's largest telescopes.
Spectroscopic resources in particular continue to be a strongly
limiting factor for such supernova surveys.  To collect much larger
numbers of these supernovae in a reasonable time period will certainly
require leaving behind operations from the ground.

\acknowledgments

We thank the staffs at all the observatories involved for their
assistance with observations.  We also thank the referee, David
Branch, for many helpful comments on the manuscript.  Partial support
for this work was provided by NASA grants GO-08641 and GO-09118 from
the Space Telescope Science Institute, which is operated by AURA,
Inc., under NASA contract NAS5-26555.  Funding was also provided by
NSF grant AST-0206329.  STH acknowledges support from the NASA LTSA
grant NAG5--9364.

\begin{deluxetable}{rlcc}
\tablewidth{0pc}
\tablecaption{Target Field Central Coordinates}
\tablehead{
\colhead{Telescope} & \colhead{Field Name} & \colhead{R.A.\tablenotemark{1}(2000)} 
& \colhead{Declination\tablenotemark{2}(2000)} 
}
\startdata
CFHT & 0230    &      02:27:53    &   +00:35:00 \\
      					 
CFHT & 0438    &      04:38:40    &  --01:30:00 \\
      					 
CFHT & 0749    &      07:49:55    &   +10:09:00 \\
      					 
CFHT & 0848    &      08:48:30    &   +44:15:00 \\
      					 
CFHT & 1052    &      10:52:20    &   +57:20:00 \\
\\
Subaru &  0230E    &      02:28:46    &   +00:35:00\\
Subaru &  0230W    &      02:27:00    &   +00:35:00\\

Subaru &  0438E    &      04:39:33    &   --01:30:00\\
Subaru &  0438W    &      04:37:47    &   --01:30:00\\

Subaru &  0749E    &      07:50:48    &   +10:09:00\\
Subaru &  0749W    &      07:49:02    &   +10:09:00\\

Subaru &  0848E    &      08:49:43    &   +44:15:00\\
Subaru &  0848W    &      08:47:17    &   +44:15:00\\

Subaru &  1052N    &      10:52:20    &   +57:33:09\\
Subaru &  1052S    &      10:52:20    &   +57:06:51\\

\enddata
\tablenotetext{1} {hh:mm:ss}
\tablenotetext{2} {dd:mm:ss}
\label{table:coords}
\end{deluxetable}

\begin{deluxetable}{ccccc}
\tablewidth{0pc}
\tablecaption{CFHT+12K Science Observations}
\tablehead{
\colhead{date} & \colhead{MJD} & \colhead{field} &
\colhead{I\tablenotemark{1}} &\colhead{FWHM\tablenotemark{2}}}
\startdata
Sep 11 &  52163.54  & 0230   &  8$\times$600     & 0.72 \\
Sep 24 &  52174.56  & 0230   &  8$\times$600     & 0.68 \\
Oct  9 &  52191.43  & 0230   &  8$\times$600     & 0.69 \\
Nov 12 &  52225.31  & 0230   &  8$\times$600     & 0.66 \\
Dec 18 &  52263.23  & 0230   &  10$\times$600    & 1.06 \\ 
total  &            & 0230   &  25200s     &  \\
\\    	    	           
Sep 12 &  52164.57  & 0438   &  8$\times$600     & 0.75 \\
Sep 24 &  52176.49  & 0438   &  8$\times$600     & 0.72 \\
Oct  9 &  52191.50  & 0438   &  8$\times$600     & 0.71 \\
Nov 12 &  52225.39  & 0438   &  8$\times$600     & 0.62 \\
Dec 18 &  52263.38  & 0438   &  8$\times$600     & 1.14 \\ 
total  &            & 0438   &  24000s     &  \\
\\    	    	           
Oct 13 &  52195.57  & 0749   &  8$\times$600     & 0.77 \\
Nov 12 &  52225.48  & 0749   &  8$\times$600     & 0.77 \\
Dec 18 &  52262.50  & 0749   &  8$\times$600     & 0.80 \\ 
total  &            & 0749   &  14400s     &  \\
\\    	    	           
Oct  9 &  52191.56  & 0848   &  8$\times$600     & 0.78 \\               
Nov 12 &  52225.56  & 0848   &  8$\times$600     & 0.77 \\               
Dec 18 &  52263.51  & 0848   &  8$\times$600     & 1.42 \\               
total  &            & 0848   &  14400s     &  \\
\\    	    	           
Dec 18 &  52263.57  & 1052   &  8$\times$600     & 1.13 \\     
total  &            & 1052   &  4800s      &  \\

\enddata
\tablenotetext{1} {All exposure times in seconds.}
\tablenotetext{2} {arcseconds}
\label{table:cfhobservations}
\end{deluxetable}

\begin{deluxetable}{ccccccccccc}
\tablewidth{0pc}
\tablecaption{Suprime-Cam Science Observations}
\tablehead{
\colhead{date} & \colhead{MJD} & \colhead{field} &
\colhead{R\tablenotemark{1}}  & \colhead{I\tablenotemark{1}} & \colhead{Z\tablenotemark{1}}& \colhead{field} &
\colhead{R\tablenotemark{1}}  & \colhead{I\tablenotemark{1}} & \colhead{Z\tablenotemark{1}}& \colhead{FWHM\tablenotemark{2}}}
\startdata
Oct 16  & 52198.47  & 0230E &  ...    & 3$\times$215 & 5$\times$240   & 0230W &  ...    & 3$\times$215 & 5$\times$240 & 0.64 \\
Oct 22  & 52204.46  & 0230E & 2$\times$280   & 3$\times$215 & 5$\times$240   & 0230W & 2$\times$280   & 3$\times$215 & 5$\times$240 & 0.66 \\
Nov 18  & 52231.43  & 0230E & 2$\times$280   & 3$\times$215 & 5$\times$240   & 0230W & 2$\times$280   & 3$\times$215 & 5$\times$240 & 0.63 \\
Nov 19  & 52232.26  & 0230E & 3$\times$280   & 3$\times$215 & 5$\times$240   & 0230W & 2$\times$280   & 3$\times$215 & 5$\times$240 & 0.64 \\
Nov 23  & 52236.24  & 0230E & 2$\times$280   & 3$\times$215 & 4$\times$240   & 0230W & 2$\times$280   & 3$\times$215 & 4$\times$240 & 0.95 \\
Dec 9   & 52252.24  & 0230E & 2$\times$280   & 3$\times$215 & 5$\times$240   & 0230W & 2$\times$280   & 3$\times$215 & 5$\times$240 & 0.94 \\
Jan 9   & 52283.31  & 0230E & 2$\times$280   & 3$\times$215 & 5$\times$240   & 0230W & 2$\times$280   & 3$\times$215 & 4$\times$240 & 0.60 \\
Jan 14  & 52288.30  & 0230E & 2$\times$280   & 3$\times$215 & 5$\times$240   & 0230W & 2$\times$280   & 3$\times$215 & 5$\times$240 & 0.70 \\
\enddata
\tablenotetext{1} {All exposure times in seconds.}
\tablenotetext{2} {arcseconds}
\label{table:subaruobservations}
\end{deluxetable}

\begin{deluxetable}{lccc}
\tablewidth{0pc}
\tablecaption{IfA Deep Survey Type Ia Supernovae}
\tablehead{
\colhead{SN} & \colhead{R.A.\tablenotemark{1} (2000)} & \colhead{Dec.\tablenotemark{2} (2000)} 
& \colhead{$E(B-V)\tablenotemark{3}$} }
\startdata
2001fo  &  04:37:41.45  & --01:29:33.1  &  0.037   \\  
2001fs  &  04:39:30.68  & --01:28:21.9  &  0.030   \\  
2001hs  &  04:39:22.39  & --01:32:51.4  &  0.032   \\  
2001hu  &  07:50:35.90  &  +09:58:14.2  &  0.021   \\  
2001hx  &  08:49:24.61  &  +44:02:22.4  &  0.031   \\  
2001hy  &  08:49:45.85  &  +44:15:31.8  &  0.032   \\  
2001iv  &  07:50:13.53  &  +10:17:10.4  &  0.022   \\  
2001iw  &  07:50:39.32  &  +10:20:19.1  &  0.022   \\
\enddata
\tablenotetext{1} {hh:mm:ss}
\tablenotetext{2} {dd:mm:ss}
\tablenotetext{3} {Galactic Extinction (mag) from Schlegel, Finkbeiner, \& Davis (1998)}
\label{table:sninfo}
\end{deluxetable}

\begin{deluxetable}{crrlr}
\tablewidth{0pc}
\tablecaption{IfA Deep Survey SNe Ia Spectroscopic Observations}
\tablehead{
\colhead{SN} & \colhead{MJD} & \colhead{Date\tablenotemark{1}} & \colhead{Instrument} & 
\colhead{Integration (s)}}
\startdata
2001fo & 52204.0  & Oct 21  & Keck-II ESI & 4400 \\  
2001fs & 52205.0  & Oct 22  & Keck-II ESI & 5400 \\  
2001hs & 52232.9  & Nov 18  & Keck-II ESI & 5400 \\  
2001hu & 52233.0  & Nov 18  & Keck-II ESI & 7800 \\  
2001hx & 52233.0  & Nov 18  & Keck-II ESI & 3000 \\  
2001hy & 52232.1  & Nov 17  & Keck-II ESI & 5500 \\  
2001iv & 52257.4  & Dec 13  &  VLT FORS1  & 2400 \\  
2001iw & 52257.3  & Dec 13  &  VLT FORS1  & 2400 \\
\enddata
\tablenotetext{1} {2001 except for 2001jn, which was taken in 2002.}
\label{table:spectralist}
\end{deluxetable}

\begin{deluxetable}{ll}
\tablewidth{0pc}
\tablecaption{SN Spectroscopic Identifications}
\tablehead{
\colhead{SN} & \colhead{Identification}}
\startdata
 2001fo & $z=0.772 \pm 0.001$ from host {\scshape [O ii]}.               	   \\  
 2001fs & $z=0.874 \pm 0.010$ from correlating with SN Ia features.   \\  
 2001hs & $z=0.833 \pm 0.001$ from host {\scshape [O ii]}, {\scshape [O iii]}.	     \\  
 2001hu & $z=0.882 \pm 0.001$ from host {\scshape [O ii]}.  Clear SN Ia features,      \\  
        & strong correlation via SNID.             \\	                
 2001hx & $z=0.799 \pm 0.001$ from host {\scshape [O ii]}.                             \\  
 2001hy & $z=0.812 \pm 0.001$ from host {\scshape [O ii]}, possible H$\beta$.          \\  
 2001iv & $z=0.3965 \pm 0.0003$ from host H$\alpha$, H$\beta$, {\scshape [O iii]}. \\
\enddata
\label{table:spectraid}
\end{deluxetable}

\begin{deluxetable}{llrcrrccr}
\tablewidth{0pc}
\tablecaption{SNID Information}
\tablehead{
\colhead{SN} & \colhead{z$_{em}\tablenotemark{1}$} & \colhead{epoch\tablenotemark{2}} & 
\colhead{z$_{best,SNID}\tablenotemark{3}$} & \colhead{r*lap\tablenotemark{3}} & 
\colhead{t$_{SNID}$\tablenotemark{3}} & 
\colhead{z$_{SNID}$(z$_{em}$)\tablenotemark{4}} & \colhead{r*lap(z$_{em}$)\tablenotemark{4}} & 
\colhead{t$_{SNID}$(z$_{em}$)\tablenotemark{4}}}
\startdata
 2001hu$^{5}$  &  0.882  & -2.9  & 0.869   &  5.8   &   -1.8  &         &       &         \\\ 
 2001iv$^{5}$  &  0.3965 &  4.2  & 0.398   &  6.7   &   -0.5  &         &       &         \\\ 
 2001iw$^{5}$  &  0.3396 &  5.1  & 0.342   &  9.3   &    2.4  &         &       &         \\\ 
 2001iy$^{5}$  &  0.568  &  9.9  & 0.570   &  6.0   &    3.8  &         &       &         \\\ 
 2001jh$^{5}$  &  0.885  &  3.5  & 0.901   &  7.3   &   -3.8  &         &       &         \\\ 
 2001jp$^{5}$  &  0.528  & -3.6  & 0.535   &  6.7   &   -5.7  &         &       &         \\\ 
 2002P$^{5}$   &  0.719  & -6.2  & 0.717   &  4.9   &   -2.8  &         &       &         \\\ 
 2002ab$^{5}$  &  0.423  &  3.9  & 0.420   &  6.8   &    0.6  &         &       &         \\
\enddata
\tablenotetext{1} {Host galaxy redshift as reported in Table ~\ref{table:spectraid}.}
\tablenotetext{2} {As determined by MLCS fits (see Section ~\ref{sec-mlcs}).}
\tablenotetext{3} {Best fit parameters as determined by SNID.  For SN
2001jf, there were no template matches with $r*lap>3.0$.  SN 2001jn
was observed in 2002, long after the SN had faded from view.}
\tablenotetext{4} {SNID parameters determined while constraining the
redshift to the host-galaxy redshift as determined by emission lines.}
\tablenotetext{5} {9 SNe for which there is unambiguous spectral
confirmation by SNID as a SN Ia at the host-galaxy redshift. }
\tablenotetext{6} {11 SNe for which there is substantial photometric
evidence, but not unambiguous spectral confirmation, for
identification as a SN Ia at the host galaxy redshift. }
\tablenotetext{7} {3 SNe for which there is clear spectral
identification by SNID as a SN Ia, but no host galaxy emission with which to
confirm the redshift. }
\label{table:snidinfo}
\end{deluxetable}

\begin{deluxetable}{cccrc}
\tablewidth{0pc}
\tablecaption{SN 2001fo observations}
\tablehead{
\colhead{MJD} & \colhead{flux\tablenotemark{1}} & \colhead{$\sigma(flux)$\tablenotemark{2}} & \colhead{epoch (rest)} & \colhead{K-correction($\sigma$)}}
\startdata
	   &	  $I$  &	  &	     & 	          $I \rightarrow B$ \\
  52164.57 &  0.000e+00 &   3.856e--02   &    -25.9 &   			\\
  52176.49 &  2.121e--02 &   1.845e--01   &    -19.2 &   			\\
  52191.50 &  2.818e+00 &   1.967e--01   &    -10.7 &   -1.128(0.040)	\\
  52204.59 &  7.106e+00 &   3.066e--01   &     -3.4 &   -1.160(0.040)	\\
  52225.39 &  5.854e+00 &   1.941e--01   &      8.4 &   -1.205(0.040)	\\
  52231.38 &  4.173e+00 &   2.044e--01   &     11.8 &   -1.235(0.040)	\\
  52232.35 &  4.051e+00 &   2.166e--01   &     12.3 &   -1.239(0.040)	\\
\enddata					   				
\tablenotetext{1} {Magnitudes may be calculated from all flux values by $mag=-2.5 \log(flux) + 25.0$ }
\tablenotetext{2} {Magnitude uncertainties may be calculated from all
flux uncertainties by $\sigma(mag) = 2.5 \log ( 1 + \sigma(flux)/flux )$.}
\label{table:baryonyx-observations}
\end{deluxetable}

\clearpage
\addtocounter{table}{-1}

\begin{deluxetable}{cccrc}
\tablewidth{0pc}
\tablecaption{SN 2001fs observations}
\tablehead{
\colhead{MJD} & \colhead{flux\tablenotemark{1}} & \colhead{$\sigma(flux)$\tablenotemark{2}} & \colhead{epoch (rest)} & 
\colhead{K-correction($\sigma$)}}
\startdata
	   &	$I$    &	    &	     &   $I \rightarrow B$ \\
  52164.57 & -3.597e--01 &  1.022e--01 &    -22.4 &                  \\
  52176.49 & -2.352e--03 &  1.819e--01 &    -16.1 &                  \\
  52191.50 &  1.617e+00 &  6.363e--02 &     -8.0 &    -1.250(0.044) \\
  52198.59 & -3.404e--01 &  7.314e--01 &     -4.3 &                  \\
  52204.58 &  2.377e+00 &  2.821e--01 &     -1.1 &    -1.243(0.044) \\
  52225.39 &  1.193e+00 &  8.805e--02 &     10.0 &    -1.215(0.044) \\
  52231.37 &  1.313e+00 &  1.819e--01 &     13.2 &    -1.176(0.044) \\
\enddata
\tablenotetext{1} {Magnitudes may be calculated from all flux values by $mag=-2.5 \log(flux) + 25.0$ }
\tablenotetext{2} {Magnitude uncertainties may be calculated from all
flux uncertainties by $\sigma(mag) = 2.5 \log ( 1 + \sigma(flux)/flux )$.}
\label{table:valdosaurus-observations}
\end{deluxetable}
      
\clearpage 
\addtocounter{table}{-1}

\begin{deluxetable}{cccrc}
\tablewidth{0pc}
\tablecaption{SN 2001hs observations}
\tablehead{
\colhead{MJD} & \colhead{flux\tablenotemark{1}} & \colhead{$\sigma(flux)$\tablenotemark{2}} & \colhead{epoch (rest)} & \colhead{K-correction($\sigma$)}}
\startdata
	   &	  $I$  &	  &	      	       &   $I \rightarrow B$ \\
  52164.57 &   6.764e--01 &   1.343e--01 &  -43.0 &       		\\
  52176.49 &   5.318e--01 &   9.576e--02 &  -36.5 &       		\\
  52191.50 &   4.296e--01 &   2.397e--01 &  -28.4 &       		\\
  52198.59 &  -1.353e+00 &   1.204e+00 &  -24.5 &       		\\
  52204.58 &   2.408e--02 &   2.069e--01 &  -21.2 &       		\\
  52225.39 &   2.430e+00 &   1.144e--01 &   -9.9 &      -1.209(0.014) 	\\
  52231.37 &   3.639e+00 &   1.215e--01 &   -6.6 &      -1.222(0.014) 	\\
\enddata					   
\tablenotetext{1} {Magnitudes may be calculated from all flux values by $mag=-2.5 \log(flux) + 25.0$ }
\tablenotetext{2} {Magnitude uncertainties may be calculated from all
flux uncertainties by $\sigma(mag) = 2.5 \log ( 1 + \sigma(flux)/flux )$.}
\label{table:archaeopteryx-observations}
\end{deluxetable}

\clearpage 
\addtocounter{table}{-1}

\begin{deluxetable}{cccrc}
\tablewidth{0pc}
\tablecaption{SN 2001hu observations}
\tablehead{
\colhead{MJD} & \colhead{flux\tablenotemark{1}} & \colhead{$\sigma(flux)$\tablenotemark{2}} & \colhead{epoch (rest)} & \colhead{K-correction($\sigma$)}}
\startdata
	   &	$I$    &	    &	      		 & $I \rightarrow B$ \\
  52195.57 &   0.000e+00 &   3.748e--01   &  -22.8 & 		  \\
  52225.48 &   2.899e+00 &   2.171e--01   &   -6.9 &    -1.258(0.034) \\   
  52231.47 &   3.554e+00 &   3.374e--01   &   -3.7 &    -1.257(0.034) \\   
  52232.45 &   3.412e+00 &   3.539e--01   &   -3.2 &    -1.257(0.034) \\   
  52252.42 &   2.157e+00 &   4.514e--01   &    7.4 &    -1.254(0.034) \\   
  52262.50 &   1.275e+00 &   4.049e--01   &   12.8 &    -1.202(0.034) \\   
  52288.50 &   1.924e--01 &   4.122e--01   &   26.6 &    -1.168(0.034) \\
\enddata
\tablenotetext{1} {Magnitudes may be calculated from all flux values by $mag=-2.5 \log(flux) + 25.0$ }
\tablenotetext{2} {Magnitude uncertainties may be calculated from all
flux uncertainties by $\sigma(mag) = 2.5 \log ( 1 + \sigma(flux)/flux )$.}
\label{table:carnotaurus-observations}
\end{deluxetable}

\clearpage 
\addtocounter{table}{-1}

\begin{deluxetable}{cccrc}
\tablewidth{0pc}
\tablecaption{SN 2001hx observations}
\tablehead{
\colhead{MJD} & \colhead{flux\tablenotemark{1}} & \colhead{$\sigma(flux)$\tablenotemark{2}} & \colhead{epoch (rest)} & \colhead{K-correction($\sigma$)}}
\startdata
	   &	  $I$   &	    & 	      &   $I \rightarrow B$ \\
  52191.56 & -2.000e--01 & 4.762e--01 &  -23.3 &			\\
  52198.61 & -8.001e--01 & 3.593e--01 &  -19.4 &			\\
  52204.61 &  2.814e--01 & 1.575e--01 &  -16.1 &     	-1.109(0.031)	\\
  52225.56 &  3.053e+00 & 6.549e--01 &   -4.4 &        -1.205(0.031) 	\\
  52231.63 &  3.689e+00 & 2.443e--01 &   -1.1 &        -1.197(0.031) 	\\
  52232.52 &  3.815e+00 & 1.257e--01 &   -0.6 &        -1.196(0.031) 	\\
  52236.50 &  3.418e+00 & 5.841e--02 &    1.6 &        -1.199(0.031) 	\\
\enddata					   				
\tablenotetext{1} {Magnitudes may be calculated from all flux values by $mag=-2.5 \log(flux) + 25.0$ }
\tablenotetext{2} {Magnitude uncertainties may be calculated from all
flux uncertainties by $\sigma(mag) = 2.5 \log ( 1 + \sigma(flux)/flux )$.}
\label{table:fabrosaurus-observations}						
\end{deluxetable}	
							
\clearpage 
\addtocounter{table}{-1}										
										
\begin{deluxetable}{cccrc}
\tablewidth{0pc}
\tablecaption{SN 2001hy observations}
\tablehead{
\colhead{MJD} & \colhead{flux\tablenotemark{1}} & \colhead{$\sigma(flux)$\tablenotemark{2}} & \colhead{epoch (rest)} & \colhead{K-correction($\sigma$)}}
\startdata
	   &	  $I$  &	  &	      	       &   $I \rightarrow B$ \\
  52191.56 &  -5.605e--02 &   3.965e--01 &      -23.7 &			\\
  52198.61 &  -2.720e--01 &   5.771e--01 &      -19.8 &			\\
  52204.61 &   8.839e--01 &   2.160e--01 &      -16.5 &  -1.166(0.023)	\\
  52225.56 &   2.976e+00 &   2.584e--01 &       -5.0 &   -1.205(0.023) \\
  52231.63 &   3.235e+00 &   1.965e--01 &       -1.6 &   -1.216(0.023) \\
  52232.52 &   3.537e+00 &   9.913e--02 &       -1.1 &   -1.217(0.023) \\
  52236.50 &   3.220e+00 &   1.682e--01 &        1.1 &   -1.222(0.023) \\
\enddata					   
\tablenotetext{1} {Magnitudes may be calculated from all flux values by $mag=-2.5 \log(flux) + 25.0$ }
\tablenotetext{2} {Magnitude uncertainties may be calculated from all
flux uncertainties by $\sigma(mag) = 2.5 \log ( 1 + \sigma(flux)/flux )$.}
\label{table:stegosaurus-observations}
\end{deluxetable}

\clearpage 
\addtocounter{table}{-1}
      
\begin{deluxetable}{cccrc}
\tablewidth{0pc}
\tablecaption{SN 2001iv observations}
\tablehead{
\colhead{MJD} & \colhead{flux\tablenotemark{1}} & \colhead{$\sigma(flux)$\tablenotemark{2}} & \colhead{epoch (rest)} & \colhead{K-correction($\sigma$)}}
\startdata
	   &	$I$    &	    &	        &  $I \rightarrow V$ \\
  52195.57 &   0.000e+00 &   2.225e--01       &  -40.1  &      	          \\    
  52225.48 &  -9.193e--01 &   4.660e--01       &  -18.7  &     	          \\    
  52231.48 &   8.874e--01 &   2.043e--01       &  -14.4  &   -0.756(0.052) \\    
  52232.46 &   1.969e+00 &   1.569e--01       &  -13.7  &   -0.757(0.052) \\    
  52236.49 &   6.649e+00 &   1.751e--01       &  -10.8  &   -0.768(0.052) \\    
  52252.43 &   2.106e+01 &   3.438e--01       &    0.6  &   -0.820(0.052) \\    
  52262.50 &   1.818e+01 &   2.727e--01       &    7.8  &   -0.767(0.052) \\
\enddata    	       
\tablenotetext{1} {Magnitudes may be calculated from all flux values by $mag=-2.5 \log(flux) + 25.0$ }
\tablenotetext{2} {Magnitude uncertainties may be calculated from all
flux uncertainties by $\sigma(mag) = 2.5 \log ( 1 + \sigma(flux)/flux )$.}
\label{table:avimimus-observations}
\end{deluxetable}      

\clearpage 
\addtocounter{table}{-1}

\begin{deluxetable}{cccrc}
\tablewidth{0pc}
\tablecaption{SN 2001iw observations}
\tablehead{
\colhead{MJD} & \colhead{flux\tablenotemark{1}} & \colhead{$\sigma(flux)$\tablenotemark{2}} & \colhead{epoch (rest)} & \colhead{K-correction($\sigma$)}}
\startdata
	   &	$I$      &	       &                  & $I \rightarrow V$ \\
  52195.57 &  -2.660e--02 &   1.514e--01 &   -41.0 &                 \\
  52225.48 &   2.826e--01 &   4.934e--01 &   -18.7 &   -0.777(0.094) \\ 
  52231.47 &   4.437e+00 &   2.353e--01 &   -14.2 &   -0.772(0.094) \\ 
  52232.45 &   5.048e+00 &   3.475e--01 &   -13.5 &   -0.772(0.094) \\ 
  52236.48 &   1.202e+01 &   1.979e--01 &   -10.5 &   -0.773(0.094) \\ 
  52252.42 &   2.971e+01 &   3.037e--01 &     1.4 &   -0.804(0.094) \\ 
  52262.50 &   1.815e+01 &   2.289e--01 &     9.0 &   -0.649(0.094) \\
\enddata
\tablenotetext{1} {Magnitudes may be calculated from all flux values by $mag=-2.5 \log(flux) + 25.0$ }
\tablenotetext{2} {Magnitude uncertainties may be calculated from all
flux uncertainties by $\sigma(mag) = 2.5 \log ( 1 + \sigma(flux)/flux )$.}
\label{table:dromaeosaurus-observations}
\end{deluxetable}

\clearpage 
\addtocounter{table}{-1}

\begin{deluxetable}{cccrc}
\tablewidth{0pc}
\tablecaption{SN 2001ix observations}
\tablehead{
\colhead{MJD} & \colhead{flux\tablenotemark{1}} & \colhead{$\sigma(flux)$\tablenotemark{2}} & \colhead{epoch (rest)} & \colhead{K-correction($\sigma$)}}
\startdata
	   &	$I$    &	   &	      	 & $I \rightarrow B$ \\
  52232.56 &  0.000e+00 &   2.125e--01 &   -21.0 &       	   \\    
  52236.65 & -4.059e--01 &   5.740e--01 &   -18.6 &       	   \\    
  52252.62 &  3.895e+00 &   3.580e--01 &    -9.3 &   -1.119(0.068) \\    
  52263.57 &  6.942e+00 &   3.101e--01 &    -2.9 &   -1.113(0.068) \\    
  52283.64 &  5.446e+00 &   3.597e--01 &     8.8 &   -1.141(0.068) \\    
  52288.63 &  4.363e+00 &   9.581e--02 &    11.7 &   -1.172(0.068) \\    
  52289.66 &  4.406e+00 &   2.265e--01 &    12.3 &   -1.179(0.068) \\
\enddata
\tablenotetext{1} {Magnitudes may be calculated from all flux values by $mag=-2.5 \log(flux) + 25.0$ }
\tablenotetext{2} {Magnitude uncertainties may be calculated from all
flux uncertainties by $\sigma(mag) = 2.5 \log ( 1 + \sigma(flux)/flux )$.}
\label{table:ozraptor-observations}
\end{deluxetable}

\clearpage 
\addtocounter{table}{-1}

\begin{deluxetable}{cccrc}
\tablewidth{0pc}
\tablecaption{SN 2001iy observations}
\tablehead{
\colhead{MJD} & \colhead{flux\tablenotemark{1}} & \colhead{$\sigma(flux)$\tablenotemark{2}} & \colhead{epoch (rest)} & \colhead{K-correction($\sigma$)}}
\startdata
	   &	$I$    &	    &		   &  $I \rightarrow V$ \\
  52232.56 &  1.446e+00 &   2.090e--02  &  -12.1 &    -0.850(0.075) \\    
  52236.65 &  4.244e+00 &   1.403e+00  &   -9.5 &    -0.879(0.075) \\    
  52252.62 &  1.085e+01 &   4.390e--01  &    0.7 &    -0.858(0.075) \\    
  52263.57 &  9.823e+00 &   3.911e--01  &    7.7 &    -0.894(0.075) \\    
  52283.64 &  4.604e+00 &   5.731e--01  &   20.5 &    -0.785(0.075) \\    
  52288.63 &  3.529e+00 &   3.423e--01  &   23.6 &    -0.765(0.075) \\    
  52289.66 &  3.781e+00 &   5.069e--01  &   24.3 &    -0.762(0.075) \\
\enddata
\tablenotetext{1} {Magnitudes may be calculated from all flux values by $mag=-2.5 \log(flux) + 25.0$ }
\tablenotetext{2} {Magnitude uncertainties may be calculated from all
flux uncertainties by $\sigma(mag) = 2.5 \log ( 1 + \sigma(flux)/flux )$.}
\label{table:majungatholus-observations}
\end{deluxetable}

\clearpage 
\addtocounter{table}{-1}

\begin{deluxetable}{cccrc}
\tablewidth{0pc}
\tablecaption{SN 2001jb observations}
\tablehead{
\colhead{MJD} & \colhead{flux\tablenotemark{1}} & \colhead{$\sigma(flux)$\tablenotemark{2}} & \colhead{epoch (rest)} & \colhead{K-correction($\sigma$)}}
\startdata
	   &	$I$    &	    &	   	&  $I \rightarrow V$ \\
  52163.54 &   3.590e--01  &  9.767e--01  &   -52.4 & 		   \\
  52174.56 &  -8.964e--02  &  2.353e--01  &   -45.9 & 		   \\
  52191.43 &   2.275e--01  &  9.035e--01  &   -36.0 & 		   \\
  52198.45 &  -3.927e--02  &  5.627e--01  &   -31.9 & 		   \\
  52204.46 &   1.756e--01  &  4.148e--01  &   -28.3 & 		   \\
  52225.31 &   3.417e--01  &  4.581e--01  &   -16.0 &  -0.819(0.150) \\
  52231.24 &   9.052e--01  &  3.463e--01  &   -12.5 &  -0.854(0.150) \\
\enddata									
\tablenotetext{1} {Magnitudes may be calculated from all flux values by $mag=-2.5 \log(flux) + 25.0$ }
\tablenotetext{2} {Magnitude uncertainties may be calculated from all
flux uncertainties by $\sigma(mag) = 2.5 \log ( 1 + \sigma(flux)/flux )$.}
\label{table:kentrosaurus-observations}						
\end{deluxetable}								
					
\clearpage 
\addtocounter{table}{-1}
									
\begin{deluxetable}{cccrc}							
\tablewidth{0pc}								
\tablecaption{SN 2001jf observations}						
\tablehead{									
\colhead{MJD} & \colhead{flux\tablenotemark{1}} & \colhead{$\sigma(flux)$\tablenotemark{2}} & \colhead{epoch (rest)} & \colhead{K-correction($\sigma$)}}
\startdata
	   &	$I$    &	    &	     		& $I \rightarrow B$ \\
  52163.54 &  -1.126e--01  &  6.847e--02   & -56.7  &			\\
  52174.56 &   3.321e--01  &  8.107e--02   & -50.6  &    		\\
  52191.43 &  -2.826e--01  &  2.078e--01   & -41.3  &			\\
  52198.45 &  -2.267e--01  &  3.849e--01   & -37.5  &			\\
  52204.46 &  -2.440e--01  &  1.220e--01   & -34.2  &			\\
  52225.31 &  -1.891e--02  &  1.850e--01   & -22.6  &			\\
  52231.24 &  -1.700e--01  &  3.864e--01   & -19.4  &			\\
\enddata								       
\tablenotetext{1} {Magnitudes may be calculated from all flux values by $mag=-2.5 \log(flux) + 25.0$ }
\tablenotetext{2} {Magnitude uncertainties may be calculated from all
flux uncertainties by $\sigma(mag) = 2.5 \log ( 1 + \sigma(flux)/flux )$.}
\label{table:massospondylus-observations}				       
\end{deluxetable}							       
				
\clearpage 
\addtocounter{table}{-1}

\begin{deluxetable}{cccrc}
\tablewidth{0pc}
\tablecaption{SN 2001jh observations}
\tablehead{
\colhead{MJD} & \colhead{flux\tablenotemark{1}} & \colhead{$\sigma(flux)$\tablenotemark{2}} & \colhead{epoch (rest)} & \colhead{K-correction($\sigma$)}}
\startdata
	   &	$I$    &	    &	      	     &   $I \rightarrow B$ \\
  52163.54 &    7.903e--02 &   2.723e--01  &   -46.1  &                   \\    
  52174.56 &   -9.379e--02 &   2.016e--01  &   -40.3  &                   \\    
  52191.43 &    2.100e--01 &   2.422e--01  &   -31.3  &                   \\    
  52198.43 &    1.713e--01 &   2.154e--01  &   -27.6  &                   \\    
  52204.48 &    6.856e--02 &   2.350e--01  &   -24.4  &                   \\    
  52225.31 &    9.170e--01 &   2.232e--01  &   -13.4  &     -1.256(0.035) \\    
  52231.23 &    2.506e+00 &   3.339e--01  &   -10.2  &     -1.258(0.035) \\
\enddata
\tablenotetext{1} {Magnitudes may be calculated from all flux values by $mag=-2.5 \log(flux) + 25.0$ }
\tablenotetext{2} {Magnitude uncertainties may be calculated from all
flux uncertainties by $\sigma(mag) = 2.5 \log ( 1 + \sigma(flux)/flux )$.}
\label{table:ornitholestes-observations}
\end{deluxetable}

\clearpage 
\addtocounter{table}{-1}

\begin{deluxetable}{cccrc}
\tablewidth{0pc}
\tablecaption{SN 2001jm observations}
\tablehead{
\colhead{MJD} & \colhead{flux\tablenotemark{1}} & \colhead{$\sigma(flux)$\tablenotemark{2}} & \colhead{epoch (rest)} & \colhead{K-correction($\sigma$)}}
\startdata
	   &	$I$    &	    &	      	        & $I \rightarrow B$ \\
  52164.57 &   2.282e--02 &   2.166e--01   &  -48.3 &         	\\     
  52176.49 &  -2.795e--02 &   2.397e--01   &  -42.1 &         	\\     
  52191.50 &  -3.037e--01 &   1.362e--01   &  -34.4 &         	\\     
  52198.59 &   4.888e--01 &   5.624e--01   &  -30.8 &          \\     
  52204.58 &   2.465e--01 &   2.108e--01   &  -27.7 &         	\\     
  52225.39 &  -2.066e--01 &   1.620e--01   &  -17.0 &         	\\     
  52231.37 &   5.518e--01 &   2.776e--01   &  -13.9 & -1.243(0.073) \\
\enddata			      		  
\tablenotetext{1} {Magnitudes may be calculated from all flux values by $mag=-2.5 \log(flux) + 25.0$ }
\tablenotetext{2} {Magnitude uncertainties may be calculated from all
flux uncertainties by $\sigma(mag) = 2.5 \log ( 1 + \sigma(flux)/flux )$.}
\label{table:suchomimus-observations}
\end{deluxetable}

\clearpage 
\addtocounter{table}{-1}

\begin{deluxetable}{cccrc}						       
\tablewidth{0pc}
\tablecaption{SN 2001jn observations}
\tablehead{
\colhead{MJD} & \colhead{flux\tablenotemark{1}} & \colhead{$\sigma(flux)$\tablenotemark{2}} & \colhead{epoch (rest)} & \colhead{K-correction($\sigma$)}}
\startdata
	   &	$I$    &	    &	      	   	&  $I \rightarrow V$ \\
  52198.59 &   -1.426e+00 &   7.186e--01   &   -40.0 &				\\
  52204.58 &   -6.884e--03 &   1.153e--01   &   -36.4 &				\\
  52231.37 &   -1.488e--01 &   3.549e--02   &   -20.1 &				\\
  52232.36 &    4.278e--01 &   8.872e--02   &   -19.5 &    			\\
  52236.33 &    8.714e--01 &   3.282e--01   &   -17.0 &      -0.790(0.136)	\\
  52252.34 &    3.196e+00 &   2.839e--01   &    -7.3 &      -0.929(0.136) 	\\
  52283.38 &    2.894e+00 &   1.419e--01   &    11.6 &      -0.797(0.136) 	\\
\enddata	       		       							
\tablenotetext{1} {Magnitudes may be calculated from all flux values by $mag=-2.5 \log(flux) + 25.0$ }
\tablenotetext{2} {Magnitude uncertainties may be calculated from all
flux uncertainties by $\sigma(mag) = 2.5 \log ( 1 + \sigma(flux)/flux )$.}
\label{table:wuerhosaurus-observations}							
\end{deluxetable}									
											
\clearpage 
\addtocounter{table}{-1}		
									
\begin{deluxetable}{cccrc}
\tablewidth{0pc}
\tablecaption{SN 2001jp observations}
\tablehead{
\colhead{MJD} & \colhead{flux\tablenotemark{1}} & \colhead{$\sigma(flux)$\tablenotemark{2}} & \colhead{epoch (rest)} & \colhead{K-correction($\sigma$)}}
\startdata
	   &	$I$    &	    &	             & $I \rightarrow V$ \\
  52198.62 &   3.052e--02  &  4.703e--01  &  -48.4 &         	\\  
  52204.63 &  -3.076e--03  &  1.493e--01  &  -44.5 &          	\\  
  52231.61 &   4.038e--01  &  2.613e--01  &  -26.8 &         	\\  
  52232.54 &   4.150e--01  &  9.331e--02  &  -26.2 &         	\\  
  52236.51 &  -5.368e--01  &  3.658e--01  &  -23.6 &          	\\  
  52252.54 &   3.125e+00  &  1.493e--01  &  -13.1 &   -0.853(0.048) \\  
  52283.50 &   1.110e+01  &  1.418e--01  &    7.1 &   -0.877(0.048) \\
\enddata
\tablenotetext{1} {Magnitudes may be calculated from all flux values by $mag=-2.5 \log(flux) + 25.0$ }
\tablenotetext{2} {Magnitude uncertainties may be calculated from all
flux uncertainties by $\sigma(mag) = 2.5 \log ( 1 + \sigma(flux)/flux )$.}
\label{table:centrosaurus-observations}
\end{deluxetable}

\clearpage 
\addtocounter{table}{-1}

\begin{deluxetable}{cccrc}
\tablewidth{0pc}
\tablecaption{SN 2001kd observations}
\tablehead{
\colhead{MJD} & \colhead{flux\tablenotemark{1}} & \colhead{$\sigma(flux)$\tablenotemark{2}} & \colhead{epoch (rest)} & \colhead{K-correction($\sigma$)}}
\startdata
	   &	  $I$  &	  &	      	       &   $I \rightarrow B$ \\
  52195.57 &   -2.751e--02  &  2.043e--01   &   -10.3 &			\\
  52225.48 &    3.423e+00  &  3.110e--01   &     5.2 &        -1.208(0.063)	\\
  52231.47 &    2.759e+00  &  1.632e--01   &     8.2 &        -1.186(0.063)	\\
  52232.45 &    2.799e+00  &  1.870e--01   &     8.8 &        -1.181(0.063)	\\
  52236.48 &    2.211e+00  &  3.785e--01   &    10.8 &        -1.155(0.063)	\\
  52252.42 &    7.468e--01  &  2.645e--01   &    19.1 &        -1.077(0.063)	\\
  52262.50 &    4.887e--01  &  2.043e--01   &    24.3 &        -1.071(0.063)	\\
\enddata   					   				
\tablenotetext{1} {Magnitudes may be calculated from all flux values by $mag=-2.5 \log(flux) + 25.0$ }
\tablenotetext{2} {Magnitude uncertainties may be calculated from all
flux uncertainties by $\sigma(mag) = 2.5 \log ( 1 + \sigma(flux)/flux )$.}
\label{table:diplodocus-observations}						
\end{deluxetable}								
					
\clearpage 
\addtocounter{table}{-1}

\begin{deluxetable}{cccrc}
\tablewidth{0pc}
\tablecaption{SN 2002P observations}
\tablehead{
\colhead{MJD} & \colhead{flux\tablenotemark{1}} & \colhead{$\sigma(flux)$\tablenotemark{2}} & \colhead{epoch (rest)} & \colhead{K-correction($\sigma$)}}
\startdata
	   &	$I$    &	  &	  	       & $I \rightarrow B$ \\ 
  52163.54 &   0.000e+00  &  6.612e--02    &   -80.8 &       	         \\    
  52174.56 &  -2.880e--01  &  1.015e--01    &   -74.4 &       	         \\    
  52191.43 &  -2.691e--01  &  1.074e--01    &   -64.6 &       	         \\    
  52198.43 &  -1.951e--01  &  2.023e--01    &   -60.5 &       	         \\    
  52204.48 &   2.285e--01  &  2.704e--01    &   -57.0 &       	         \\    
  52225.31 &   1.538e--01  &  1.925e--01    &   -44.9 &       	         \\    
  52231.23 &   2.003e--01  &  3.201e--01    &   -41.5 &       	         \\
\enddata
\tablenotetext{1} {Magnitudes may be calculated from all flux values by $mag=-2.5 \log(flux) + 25.0$ }
\tablenotetext{2} {Magnitude uncertainties may be calculated from all
flux uncertainties by $\sigma(mag) = 2.5 \log ( 1 + \sigma(flux)/flux )$.}
\label{table:animantarx-observations}
\end{deluxetable}

\clearpage 
\addtocounter{table}{-1}

\begin{deluxetable}{cccrc}
\tablewidth{0pc}
\tablecaption{SN 2002W observations}
\tablehead{
\colhead{MJD} & \colhead{flux\tablenotemark{1}} & \colhead{$\sigma(flux)$\tablenotemark{2}} & \colhead{epoch (rest)} & \colhead{K-correction($\sigma$)}}
\startdata
	   &	  $I$  &	  &	      	       &   $I \rightarrow B$ \\
  52191.56 &  -8.985e--02  &  6.036e--01   &   -53.7  &		   \\
  52198.62 &   8.539e--02  &  5.771e--01   &   -50.2  &      	          \\    
  52204.63 &   2.058e--01  &  1.752e--01   &   -47.2  &      	          \\    
  52225.56 &  -1.201e+00  &  3.222e--01   &   -36.9  &       	          \\ 	 
  52231.61 &   2.252e--01  &  9.028e--02   &   -33.9  &      	          \\    
  52232.54 &   1.137e--01  &  2.142e--01   &   -33.5  &      	          \\    
  52236.51 &   1.473e--01  &  1.239e--01   &   -31.5  &      	          \\
\enddata					   
\tablenotetext{1} {Magnitudes may be calculated from all flux values by $mag=-2.5 \log(flux) + 25.0$ }
\tablenotetext{2} {Magnitude uncertainties may be calculated from all
flux uncertainties by $\sigma(mag) = 2.5 \log ( 1 + \sigma(flux)/flux )$.}
\label{table:afrovenator-observations}
\end{deluxetable}

\clearpage 
\addtocounter{table}{-1}

\begin{deluxetable}{cccrc}							
\tablewidth{0pc}								
\tablecaption{SN 2002X observations}
\tablehead{
\colhead{MJD} & \colhead{flux\tablenotemark{1}} & \colhead{$\sigma(flux)$\tablenotemark{2}} & \colhead{epoch (rest)} & \colhead{K-correction($\sigma$)}}
\startdata
	   &	  $I$  &	  &	   	        &   $I \rightarrow B$ \\
  52191.56 &  -5.275e--04  &  3.540e--02   & -30.1    &       \\
  52198.62 &   1.446e--01  &  2.407e--01   & -26.3    &       \\
  52204.63 &  -1.292e--02  &  1.469e--01   & -23.1    &       \\
  52225.56 &   5.081e--02  &  3.983e--01   & -11.8    &     -1.223(0.027)  \\
  52231.61 &   2.602e--02  &  1.929e--01   &  -8.6    &     -1.235(0.027)  \\
  52232.54 &   1.588e--01  &  2.460e--01   &  -8.1    &     -1.237(0.027)  \\
  52236.51 &   1.623e--01  &  2.018e--01   &  -5.9    &     -1.243(0.027)  \\
\enddata   					   
\tablenotetext{1} {Magnitudes may be calculated from all flux values by $mag=-2.5 \log(flux) + 25.0$ }
\tablenotetext{2} {Magnitude uncertainties may be calculated from all
flux uncertainties by $\sigma(mag) = 2.5 \log ( 1 + \sigma(flux)/flux )$.}
\label{table:barosaurus-observations}
\end{deluxetable}
	 
\clearpage 
\addtocounter{table}{-1}  
	   
\begin{deluxetable}{cccrc}							
\tablewidth{0pc}								
\tablecaption{SN 2002aa observations}
\tablehead{
\colhead{MJD} & \colhead{flux\tablenotemark{1}} & \colhead{$\sigma(flux)$\tablenotemark{2}} & \colhead{epoch (rest)} & \colhead{K-correction($\sigma$)}}
\startdata
	   &	  $I$  &	  &	      	       &   $I \rightarrow B$ \\
  52195.57 &    3.366e--01  &  3.694e--01   &    -40.0 &    			\\
  52225.48 &   -2.170e+00  &  4.113e--01   &    -24.6 &			\\
  52231.48 &   -3.183e--02  &  9.303e--02   &    -21.6 &			\\
  52232.46 &   -1.020e--01  &  1.952e--01   &    -21.1 &			\\
  52236.49 &   -1.775e+00  &  5.116e--01   &    -19.1 &			\\
  52252.43 &    5.336e--01  &  4.478e--01   &    -10.8 &       -1.241(0.069)	\\
  52262.50 &    3.659e+00  &  2.061e--01   &     -5.7 &       -1.243(0.069)	\\
\enddata					   				
\tablenotetext{1} {Magnitudes may be calculated from all flux values by $mag=-2.5 \log(flux) + 25.0$ }
\tablenotetext{2} {Magnitude uncertainties may be calculated from all
flux uncertainties by $\sigma(mag) = 2.5 \log ( 1 + \sigma(flux)/flux )$.}
\label{table:koparion-observations}						
\end{deluxetable}								
					
\clearpage 
\addtocounter{table}{-1}

\begin{deluxetable}{cccrc}
\tablewidth{0pc}
\tablecaption{SN 2002ab observations}
\tablehead{
\colhead{MJD} & \colhead{flux\tablenotemark{1}} & \colhead{$\sigma(flux)$\tablenotemark{2}} & \colhead{epoch (rest)} & \colhead{K-correction($\sigma$)}}
\startdata
	   &	$I$    &	    &	      	   	 & $I \rightarrow V$ \\
  52195.57 &  -3.882e--02  &  1.943e--01  &   -63.9 &		       \\
  52225.48 &   1.281e--01  &  4.897e--01  &   -42.9 &                 \\   
  52231.48 &  -1.752e+00  &  2.298e--01  &   -38.7 &                 \\   
  52232.46 &   2.813e--01  &  2.718e--01  &   -38.0 &                 \\   
  52236.49 &   1.892e--01  &  2.563e--01  &   -35.1 &                 \\   
  52252.43 &   7.209e--01  &  1.733e--01  &   -23.9 &                 \\   
  52262.50 &   2.593e--02  &  2.289e--01  &   -16.9 &   -0.789(0.048) \\
\enddata
\tablenotetext{1} {Magnitudes may be calculated from all flux values by $mag=-2.5 \log(flux) + 25.0$ }
\tablenotetext{2} {Magnitude uncertainties may be calculated from all
flux uncertainties by $\sigma(mag) = 2.5 \log ( 1 + \sigma(flux)/flux )$.}
\label{table:gargoyleosaurus-observations}
\end{deluxetable}

\clearpage 
\addtocounter{table}{-1}

\begin{deluxetable}{cccrc}
\tablewidth{0pc}
\tablecaption{SN 2002ad observations}
\tablehead{
\colhead{MJD} & \colhead{flux\tablenotemark{1}} & \colhead{$\sigma(flux)$\tablenotemark{2}} & \colhead{epoch (rest)} & \colhead{K-correction($\sigma$)}}
\startdata
	   &	$I$    &	    &	      	   	&  $I \rightarrow V$ \\
  52232.55 &   3.209e--01  &  3.876e--01    &  -24.1  &        	   \\     
  52236.63 &  -1.001e+00  &  6.846e--01    &  -21.4  &           	   \\     
  52252.61 &   4.556e+00  &  2.909e--01    &  -10.9  &    -0.863(0.037) \\     
  52263.57 &   1.123e+01  &  3.057e--01    &   -3.6  &    -0.867(0.037) \\     
  52283.63 &   9.458e+00  &  3.449e--01    &    9.6  &    -0.880(0.037) \\     
  52288.62 &   8.537e+00  &  2.865e--01    &   12.9  &    -0.882(0.037) \\     
  52323.56 &   1.694e+00  &  4.599e--01    &   36.0  &    -0.826(0.037) \\
\enddata
\tablenotetext{1} {Magnitudes may be calculated from all flux values by $m=-2.5 \log(flux) + 25.0$ }
\tablenotetext{2} {Magnitude uncertainties may be calculated from all
flux uncertainties by $\sigma(mag) = 2.5 \log ( 1 + \sigma(flux)/flux )$.}
\label{table:parvicursor-observations}
\end{deluxetable}

\clearpage 
										
\begin{deluxetable}{cccccrc}
\tablewidth{0pc}
\tablecaption{HST WFPC2 F850LP Observations}
\tablehead{
\colhead{SN} & \colhead{MJD} & \colhead{Integration (sec)} & \colhead{mag} & \colhead{$\sigma(mag)$} 
& \colhead{epoch (rest)} & \colhead{K-correction($\sigma$)}}
\startdata
 2001hu &  52280.23 &  5600.0   &   23.880  &    0.106   &   15.8  &   -0.929(0.138) \\
        &  52287.10 &  5600.0 	&   24.304  &    0.152   &   19.4  &   -0.851(0.138) \\
        &  52294.45 &  5600.0 	&   24.422  &    0.163   &   23.3  &   -0.773(0.138) \\
        &  52301.45 &  7500.0 	&   24.663  &    0.173   &   27.0  &   -0.710(0.138) \\
        &  52305.19 &  7500.0 	&   24.499  &    0.148   &   29.0  &   -0.682(0.138) \\
\\				   			     
 2001jf &  52271.07 &  5600.0   &   23.551  &     0.105  &    2.5  &   -0.864(0.104) \\
        &  52277.08 &  5600.0   &   23.615  &     0.126  &    5.9  &   -0.905(0.104) \\
        &  52284.09 &  5600.0   &   24.121  &     0.168  &    9.7  &   -0.902(0.104) \\
        &  52291.50 &  5600.0   &   24.198  &     0.173  &   13.8  &   -0.856(0.104) \\
        &  52299.32 &  7500.0   &   24.276  &     0.148  &   18.1  &   -0.794(0.104) \\
        &  52303.26 &  7500.0   &   24.301  &     0.157  &   20.3  &   -0.763(0.104) \\
\\							     
 2001jh &  52246.70 &  5600.0 	&   23.894  &    0.134   &    4.4  &   -0.975(0.136) \\
        &  52253.52 &  5600.0 	&   24.091  &    0.148   &    8.0  &   -0.968(0.136) \\
        &  52259.79 &  5600.0 	&   23.376  &    0.195   &   11.3  &   -0.885(0.136) \\
        &  52266.48 &  5600.0 	&   24.424  &    0.184   &   14.9  &   -0.776(0.136) \\
        &  52275.69 &  7500.0 	&   25.582  &    0.323   &   19.8  &   -0.671(0.136) \\
        &  52282.63 &  7500.0 	&   25.644  &    0.428   &   23.5  &   -0.629(0.136) \\
\enddata				
\label{table:hstobs}			
\end{deluxetable}

\clearpage

\begin{deluxetable}{llcccrr}
\tablewidth{0pc}
\tablecaption{SN Ia MLCS Fit Parameters}
\tablehead{
\colhead{SN} & \colhead{redshift} & \colhead{$t_{max}$} & \colhead{m-M($\sigma$)\tablenotemark{1}} &
\colhead{$A_{V}(\sigma)$} & \colhead{$\Delta$} & \colhead{$\chi^{2}$/N}}
\startdata
2001fo$^{3}$  &  0.772  & 52210.5 &   42.81(0.22) &   0.04(0.26)   &    -0.08 &  0.82   \\  
2001fs$^{4}$  &  0.874  & 52206.6 &   43.18(0.36) &   0.08(0.42)   &     0.59 &  3.91   \\  
2001hs$^{3}$  &  0.833  & 52243.5 &   43.18(0.30) &   0.01(0.25)   &     0.15 &  0.58   \\  
2001hu$^{2}$  &  0.882  & 52238.5 &   43.46(0.38) &   0.00(0.36)   &     0.30 &  1.17   \\  
2001hx$^{3}$  &  0.799  & 52233.5 &   43.65(0.36) &   0.03(0.29)   &     0.01 &  1.49   \\  
2001hy$^{3}$  &  0.812  & 52234.5 &   43.52(0.41) &   0.01(0.35)   &     0.31 &  1.11   \\  
2001iv$^{2}$  &  0.3965 & 52251.6 &   41.45(0.17) &   0.03(0.17)   &    -0.03 &  2.49   \\  
2001iw$^{2}$  &  0.3396 & 52250.5 &   41.24(0.17) &   0.18(0.16)   &    -0.28 &  3.94   \\
\enddata
\tablenotetext{1} {all distances calculated using $H_{0}=72$ km s$^{-1}$ Mpc$^{-1}$ (Freedman et al. 2001)}
\tablenotetext{2} {9 SNe for which there is unambiguous spectral
confirmation as a SN Ia at the host-galaxy redshift. }
\tablenotetext{3} {11 SNe for which there is substantial photometric
evidence, but not unambiguous spectral confirmation, for
identification as a SN Ia at the host galaxy redshift. }
\tablenotetext{4} {3 SNe for which there is clear spectral
identification as a SN Ia, but no host galaxy emission with which to
confirm the redshift. }
\label{table:mlcsinfo}
\end{deluxetable}

\clearpage

\begin{deluxetable}{llccccc}
\tablewidth{0pc}
\tablecaption{SN Ia Light Curve Fit Parameters}
\tablehead{
\colhead{SN} & \colhead{redshift} & \colhead{m-M$_{BATM}$($\sigma$)\tablenotemark{1}} & \colhead{$A_{V,BATM}(\sigma)$} & 
\colhead{m-M$_{dm15}$($\sigma$)\tablenotemark{1}} & \colhead{$E(B-V)_{dm15}(\sigma)$} & \colhead{dm15} }
\startdata
2001fo$^{3}$  &  0.772  & 43.00(0.20) & 0.04(0.08)   &   42.88(0.18)  &    0.07(0.03) &   1.08(0.12)    \\  
2001fs$^{4}$  &  0.874  & 43.84(0.50) & 0.49(0.34)   &     ...        &      ...      &      ...        \\  
2001hs$^{3}$  &  0.833  & 43.45(0.34) & 0.20(0.27)   &     ...        &      ...      &      ...        \\  
2001hu$^{2}$  &  0.882  & 43.74(0.51) & 0.40(0.34)   &     ...        &      ...      &      ...        \\  
2001hx$^{3}$  &  0.799  & 43.81(0.44) & 0.29(0.31)   &   43.24(0.33)  &    0.20(0.07) &   1.27(0.32)    \\  
2001hy$^{3}$  &  0.812  & 43.96(0.37) & 0.18(0.26)   &     ...        &      ...      &      ...        \\  
2001iv$^{2}$  &  0.3965 & 41.20(0.22) & 0.54(0.21)   &   41.31(0.25)  &    0.20(0.05) &   1.09(0.16)    \\  
2001iw$^{2}$  &  0.3396 & 40.78(0.32) & 0.58(0.22)   &   40.88(0.21)  &    0.23(0.03) &   1.13(0.20)    \\
\enddata
\tablenotetext{1} {all distances calculated using $H_{0}=72$ km s$^{-1}$ Mpc$^{-1}$ (Freedman et al. 2001)}
\tablenotetext{2} {9 SNe for which there is unambiguous spectral
confirmation as a SN Ia at the host-galaxy redshift. }
\tablenotetext{3} {11 SNe for which there is substantial photometric
evidence, but not unambiguous spectral confirmation, for
identification as a SN Ia at the host galaxy redshift. }
\tablenotetext{4} {3 SNe for which there is clear spectral
identification as a SN Ia, but no host galaxy emission with which to
confirm the redshift. }
\label{table:otherfitinfo}
\end{deluxetable}

\clearpage

\begin{deluxetable}{lcrlcccc}
\tablewidth{0pc}
\tablecaption{SN Ia Summary}
\tablehead{
\colhead{SN} & \colhead{$l^{II}$} & \colhead{$b^{II}$} & \colhead{$z$} & 
\colhead{log($cz$)} & \colhead{$\langle\log(dH_{0})\rangle$} & \colhead{$\pm$} & 
\colhead{$\langle A_{V}\rangle$} }
\startdata
2001fo$^{2}$  &  197.647 &  -30.139  &   0.772  &  5.364  &   5.434  &   0.030  &   0.10  \\
2001fs$^{3}$  &  197.895 &  -29.739  &   0.874  &  5.418  &   5.560  &   0.076  &   0.28  \\
2001hs$^{2}$  &  197.949 &  -29.806  &   0.833  &  5.397  &   5.521  &   0.057  &   0.11  \\
2001hu$^{1}$  &  210.560 &   17.621  &   0.882  &  5.422  &   5.577  &   0.078  &   0.20  \\
2001hx$^{2}$  &  176.770 &   39.195  &   0.799  &  5.379  &   5.587  &   0.060  &   0.32  \\
2001hy$^{2}$  &  176.490 &   39.264  &   0.812  &  5.386  &   5.605  &   0.069  &   0.10  \\
2001iv$^{1}$  &  210.219 &   17.674  &   0.3965 &  5.076  &   5.120  &   0.034  &   0.41  \\
2001iw$^{1}$  &  210.217 &   17.792  &   0.3396 &  5.008  &   5.034  &   0.033  &   0.50  \\
\enddata
\tablenotetext{1} {9 SNe for which there is unambiguous spectral
confirmation as a SN Ia at the host-galaxy redshift. }
\tablenotetext{2} {11 SNe for which there is substantial photometric
evidence, but not unambiguous spectral confirmation, for
identification as a SN Ia at the host galaxy redshift. }
\tablenotetext{3} {3 SNe for which there is clear spectral
identification as a SN Ia, but no host galaxy emission with which to
confirm the redshift. }
\tablenotetext{4} {SN 2001jn is not included in our cosmological fits
due to $\langle A_{V}\rangle > 0.5$.}

\label{table:summarytable}
\end{deluxetable}

\clearpage

\clearpage

\clearpage

\clearpage

\clearpage

\clearpage

\clearpage

\clearpage

\clearpage
\begin{figure}
\epsscale{2.2}
\plotone{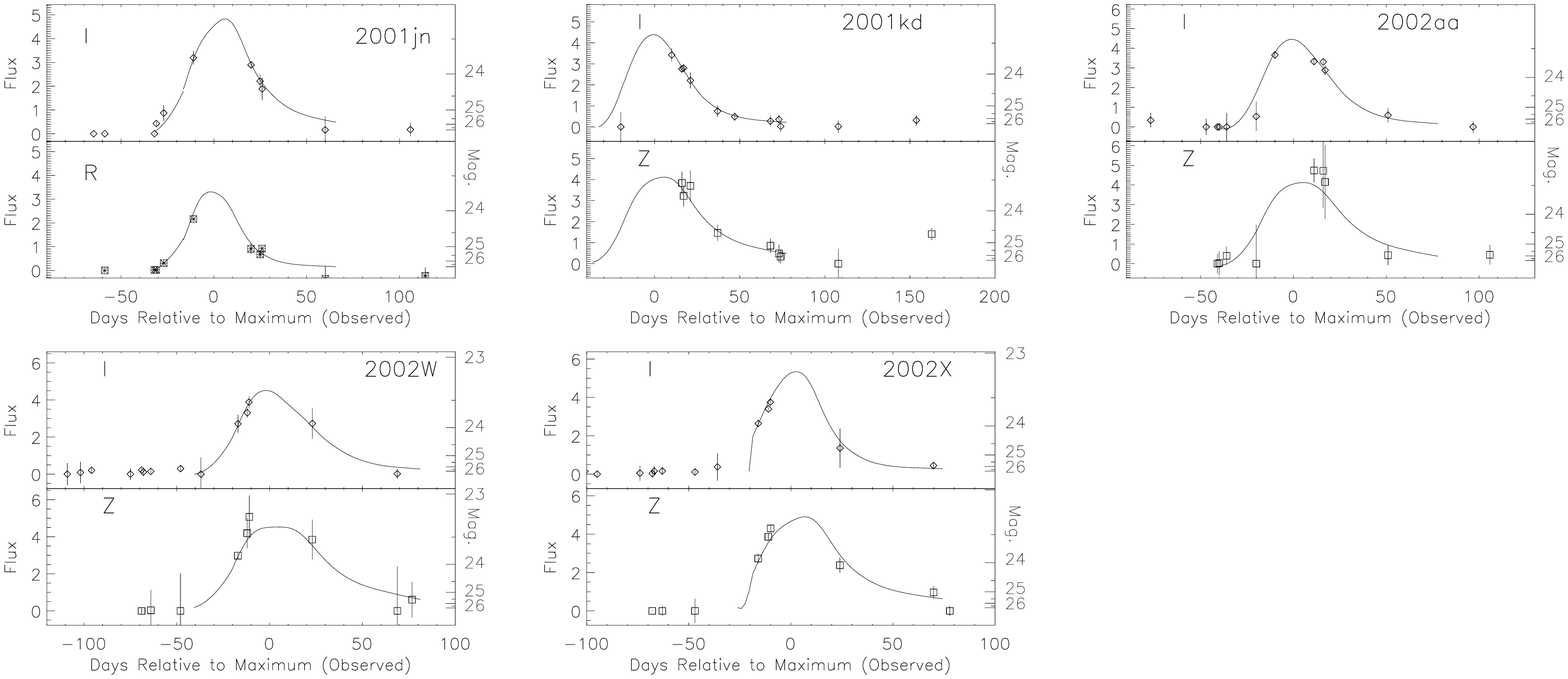}
\caption{Light curves for the 14 additional SNe Ia.
Ground-based data are plotted as observed for the two filters fit with
MLCS ($R$,$I$ for $z<0.7$ and $I$,$Z$ for $z>0.7$), with the MLCS fit
shown by solid lines.  In addition, $F850LP$ (triangles) points for SN
2001jf are shown with a shift equal to the difference in K-corrections
from $Z \rightarrow V$ and $F850LP \rightarrow V$ to better illustrate
the fit to the data. }
\label{lc-more2}
\end{figure}
\clearpage

\clearpage

\clearpage

\clearpage

\end{document}